\documentstyle[editedvolume,numreferences]{crckapb} 

\newcommand{\bra}[1]{\langle{#1}|}
\newcommand{\ket}[1]{|{#1}\rangle}

\newcommand{\eq}[1]{Eq.~(\ref{#1})}

\def\beq{\begin{equation}}
\def\eeq{\end{equation}}
\def\beqa{\begin{eqnarray}}
\def\eeqa{\end{eqnarray}}

\renewcommand{\theequation}{\thesection.\arabic{equation}}
\newcommand{\EQ}{\begin{equation}}
\newcommand{\EN}{\end{equation}}
\newcommand{\bea}{\begin{eqnarray}}
\newcommand{\ena}{\end{eqnarray}}

\renewcommand{\a}{\alpha}
\renewcommand{\b}{\beta}

\newcommand{\NP}[1]{Nucl.\ Phys.\ {\bf #1}}
\newcommand{\PL}[1]{Phys.\ Lett.\ {\bf #1}}

\newcommand{\dpb}{D$p$-brane}
\newcommand{\dpbs}{D$p$-branes}

\def\one{{\hbox{ 1\kern-.8mm l}}}

\def\sgh{{\rm sgh}}
\def\NS{{\rm NS}}
\def\R{{\rm R}}
\def\ii{{\rm i}}

\newlength{\bredde}
\def\slash#1{\settowidth{\bredde}{$#1$}\ifmmode\,\raisebox{.15ex}{/}
\hspace*{-\bredde} #1\else$\,\raisebox{.15ex}{/}\hspace*{-\bredde} #1$\fi}

\begin{opening}

\title{D branes in string theory, I}

\subtitle{Lectures presented at the 1999 NATO-ASI on ``Quantum Geometry'' 
in Akureyri, Iceland. \\ Preprint: NORDITA-1999/77 HE, hep-th/9912161. \\
Work partially supported by the 
European Commission TMR programme ERBFMRX-CT96-0045 and by 'Programma di
breve mobilit{\'{a}} per scambi internazionali dell'Universit{\'{a}} di 
Napoli "Federico II"'.}

\author{Paolo Di Vecchia}

\institute{Nordita, \\
           Blegdamsvej 17, 2100 Copenhagen OE \\
           Denmark\\
           email: divecchia@nbivms.nbi.dk}
\author{Antonella Liccardo}

\institute{Dipartimento di Fisica, Universit{\'{a}} di Napoli, \\
          and I.N.F.N. , Sezione di Napoli\\
           Mostra d'Oltremare, Pad. 19, 80125 Napoli \\
           Italy\\
           email: liccardo@na.infn.it}

\end{opening}

\runningtitle{D branes}

\begin{document}

\begin{abstract}
In these lectures we present a detailed description of the origin and of
the construction of the boundary state that is now widely used for 
studying the properties of D branes.
\end{abstract}

\section{Introduction}
\label{sec:intro}

The existence of D$p$-branes in string theories has been  an
essential ingredient for concluding that the five consistent and 
perturbatively inequivalent supersymmetric string theories in ten dimensions 
belong to a unique eleven dimensional theory that is called
M-theory. In the framework of string theories their existence was required
by T-duality in theories with both open and closed strings~\footnote{See 
Ref.~\cite{lectPOL} and references therein.}. On the other hand
classical solutions of the low-energy string effective action coupled to 
graviton, dilaton and  $(p+1)$-form  R-R potential were later 
constructed~\footnote{See Ref.~\cite{REV} and references therein.}.
Since their tension is proportional to the inverse of the string coupling 
constant they correspond to new non-perturbative states of string theory.
At the end of 1995 Polchinski~\cite{POL95} provided strong arguments for 
identifying this new  states with the D$p$-branes required by T-duality
opening the way to study their properties in string theory. In particular
their interaction can be computed through the one-loop open string 
annulus diagram. On the other hand, since the very early days of string 
theory it is known
that this one-loop open string diagram can be equivalently rewritten as a 
tree diagram in the closed string theory in which a closed string is generated
from the vacuum, propagates for a while and then annihilates again in the 
vacuum. The state that describes the creation of closed string from the 
vacuum is called the boundary state, that first appeared in the 
literature~\cite{EARLY}
in the early days of string theory for factorizing the planar and
non-planar loops of open string in the closed string channel. In the middle
of the eighties after that the BRST invariant formulation of string theory
became available the boundary states was considered again in a series of
beautiful papers by Callan et al.~\cite{CALLAN}, where, among other things, the 
ghost contribution was
added and the boundary state with an external abelian gauge field
was constructed. It was also used for deriving the gauge group of  open
string theories by requiring the tadpole cancellation~\cite{PCAI}. Its
extension to the case of Dirichlet boundary conditions was given in a series
of beautiful papers written by M.Green et al.~\cite{green} for studying 
D$p$-branes before it
became clear that they were new states of string theory corresponding to
the classical solution of the low-energy string effective action. In the
last few years the boundary state has been widely used for studying properties
of D branes in string theory.

In these lectures after a review of the main properties of perturbative string
theory we discuss in detail T-duality for both open and closed string
theories and we show how the requirement of T-duality in presence of open 
strings implies the existence of D$p$-branes that are then identified with
the new non-perturbative states obtained as classical solutions of the 
low-energy string effective action. Then, by requiring that the interaction
between two D$p$-branes gives the same result if we compute it in the 
open or in the closed string channel, we construct the boundary state that
provides a stringy description of the simplest D$p$ brane solutions. Finally
we show how to connect the boundary state to the supergravity classical
solutions.
 
\section{Perturbative String Theory}
\label{sec:secondsection}

The action of the bosonic string theory is
\beq
\label{stact}
S=-\frac{T}{2}\int_M d^2 \xi \sqrt {-h}~ h^{\alpha\beta} \partial_\alpha
X^\mu\partial_\beta X_\mu~,
\eeq
where $T$ is the string tension, $M$ is the world-sheet of the string
described by the world sheet coordinate $\xi^{\alpha} \equiv (\tau,
\sigma)
$, $h^{\alpha\beta} $  is the world sheet metric tensor
and $h={\rm det} h_{\alpha\beta}.$
The string tension is related to the Regge slope by $T=(2\pi\alpha')^{-1}$.

The  action in eq.(\ref{stact}) is invariant under local reparametrizations 
of the world sheet coordinates corresponding
to $\xi^\alpha \rightarrow f^\alpha (\xi)$ and under
local Weyl transformations  corresponding to a local rescaling of the
metric tensor  $h_{\alpha\beta} \rightarrow \Lambda(\xi) h_{\alpha\beta}.$
These symmetries allow one to bring the metric tensor in the form
$h_{\alpha\beta}=e^{\phi}\eta_{\alpha\beta}$. This choice is referred to as
the conformal gauge choice.

In this gauge the string action is still invariant under some residual local
symmetries. It is in fact invariant under a combination of a Weyl rescaling
and a local reparametrization
$\xi^\alpha\rightarrow\xi^\alpha+\varepsilon^\alpha$
($f^{\alpha} = 1+ \epsilon^{\alpha} $)
satisfying the following condition
\beq
\label{cfin}
\partial^\alpha\varepsilon^\beta+\partial^\beta\varepsilon^\alpha=
\Lambda (\sigma) \eta^{\alpha\beta},
\eeq
which corresponds to an infinitesimal conformal transformation.
String theory in the conformal gauge is then conformal invariant. 

The equation of motion for the string coordinate $X^{\mu}$ following from
the action in eq.(\ref{stact}) in the conformal gauge is given by
\beq
\label{steq}
(\partial_\sigma^2-\partial_\tau^2)X^\mu=0~~~~,~~~~ \mu = 0,..., d-1,
\eeq
while that for the metric implies the vanishing of the world sheet
energy-momentum tensor
\beq
\label{steq2}
T_{\alpha \beta} \equiv \frac{2}{T\sqrt{-h}}\frac{\delta S}{\delta
h^{\alpha\beta}}=
\partial_{\alpha} X  \cdot \partial_{\beta} X -\frac{1}{2}\eta_{\alpha
\beta}
\partial_{\gamma} X  \cdot \partial^{\gamma} X =0,
\eeq
where $d$ is the number of dimensions of the
embedding space-time. By varying the action in eq.(\ref{stact}) in the
conformal gauge, in addition to the previous eqs. of motion, we
must also impose the following boundary conditions:
\beq
\label{stbc}
\int  d\tau\left( \partial_\sigma X \cdot
\delta X |_{\sigma=\pi}- \frac{1}{2} \partial_\sigma X \cdot
\delta X |_{\sigma=0}\right)=0,
\eeq
where we have taken $\sigma\in [0,{\pi}]$

The previous boundary conditions can be satisfied in two different ways
leading to two different theories. By imposing the periodicity condition
\beq
\label{bccl}
X^\mu(\tau,0)=X^\mu(\tau,\pi),
\eeq
we obtain a closed string theory, while requiring
\beq
\label{bcop}
\partial_\sigma X_\mu\delta X^\mu|_{0,\pi}=0,
\eeq
separately at both $\sigma=0$ and $\sigma=\pi$ we obtain an open string
theory. In this latter case eq.(\ref{bcop}) can be satisfied in either of
the
two ways
\beq
\label{neudic}
\left\{ \begin{array}{l}
\partial_\sigma X_\mu|_{0,\pi}=0
\rightarrow {\rm Neumann~boundary~conditions}\\
\delta X^\mu|_{0,\pi}=0\rightarrow {\rm Dirichlet~boundary~conditions}.
\end{array}
\right.
\eeq
If the open string satisfies Neumann boundary conditions at both its
endpoints
(N-N boundary conditions) the general solution of the eqs.(\ref{steq}) and
(\ref{stbc})  is equal to
\beq
\label {expo1}
X^\mu(\tau ,\sigma)= q^\mu+2\alpha'p^\mu \tau +i\sqrt{2\alpha' }\sum_{n\neq
0}
\left(\frac {\alpha^{\mu}_{n}}{n}e^{-in \tau}\cos{n\sigma}\right),
\eeq
where $n$ is an integer. In order to have more compact expressions
without any distinction between the zero and non-zero modes it is convenient
to introduce $\alpha^\mu_0 = \sqrt{2\alpha'} p^\mu$. 
For D-D boundary conditions we have
\beq
\label {expo2}
X^\mu(\tau ,\sigma)= \frac{c^\mu(\pi-\sigma)+d^\mu \sigma}{\pi}
-\sqrt{2\alpha' }\sum_{n\neq 0}
\left(\frac {\alpha^{\mu}_{n}}{n}e^{-in \tau}\sin{n\sigma}\right).
\eeq
Finally for mixed boundary conditions we have
\beq
\label {expo3}
X^\mu(\tau ,\sigma)= c^\mu-\sqrt{2\alpha' }\sum_{r\in Z+\frac{1}{2}}
\left(\frac {\alpha^{\mu}_{r}}{r}e^{-ir \tau}\sin{r\sigma}\right),
\eeq
in the case of D-N boundary conditions and
\beq
\label {expo4}
X^\mu(\tau ,\sigma)= d^\mu+i\sqrt{2\alpha' }\sum_{r\in Z+\frac{1}{2}}
\left(\frac {\alpha^{\mu}_{r}}{r}e^{-ir \tau}\cos{r\sigma}\right),
\eeq
for N-D boundary conditions.
$c^\mu$ and $d^\mu$ are two constant vectors describing the position of
the two endpoints of the string in the embedding space-time.
Among the four solutions in eqs. (\ref{expo1})-(\ref{expo4}) the only one
which
is Poincar\'e invariant is the one correponding to N-N boundary conditions.
In the following,  unless explicitly mentioned, we will refer to this
case.

Passing to the case of a closed string the most general solution of the eqs.
of motion and of the periodicity condition in eq.(\ref{bccl}) can be written
as follows
\beq
\label {expc}
X^\mu(\tau, \sigma)= q^\mu+2\alpha'p^\mu \tau +i\sqrt{\frac{\alpha'}{2} }
\sum_{n\neq 0}
\left(\frac {\alpha^{\mu}_{n}}{n}e^{-2in(\tau -\sigma)}+
\frac {\widetilde\alpha^{\mu}_{n}}{n}e^{-2in(\tau+\sigma)}\right),
\eeq
Also here it is convenient to introduce the notation
$\alpha^\mu_0 = {\widetilde{\alpha}}^\mu_0 = p^\mu
\sqrt{\frac{\alpha'}{2}}$.

The world sheet energy-momentum tensor given in eq.(\ref{steq2}) is
conserved
if the string eqs. of motion are satisfied and is also traceless
as a consequence of the invariance under Weyl rescaling.
It is useful to rewrite it  in the light cone coordinates
\beq
\label{lccor}
\xi_+=\tau+\sigma ~~~~~~~;~~~~~~~\xi_-=\tau-\sigma ,
\eeq
where its two independent components are
\beq
\label{emtenlc}
T_{++}=\partial_+ X \cdot \partial_+ X~~~~~~~;~~~~~~~
T_{--}=\partial_- X \cdot \partial_-X.
\eeq
They are both vanishing as a consequence of the eq. of motion
for the metric in eq.(\ref{steq2}).

Inserting in the previous eqs. the mode expansion for a closed string we get
\beq
\label{liccoem}
T_{++} \sim \sum_{n\in Z}\widetilde L_{n}e ^{-2in (\tau+\sigma)}
~~~~~~~;~~~~~~~
T_{--} \sim \sum_{n\in Z} L_{n}e ^{-2in (\tau-\sigma)}~,
\eeq
where $L_{n}$ and $\widetilde L_{n}$ are given by
\beq
\label{vgenc}
L_{n}=\frac{1}{2}\sum_{m\in Z}\alpha_{-m} \cdot \alpha_{n+m}
~~~~~;~~~~~
\widetilde L_{n}=\frac{1}{2}\sum_{m\in Z}\widetilde\alpha_{-m} \cdot
\widetilde\alpha_{n+m}.
\eeq
and $\alpha_0$ and ${\tilde{\alpha}}_0$ are defined after eq.(\ref{expc})
in terms of the momentum.
In the case of an open string we have only one set of Virasoro generators:
\beq
\label{vgeno}
L_{n}=\frac{1}{2}\sum_{m\in Z}\alpha_{-m} \cdot \alpha_{n+m}~.
\eeq
where $\alpha_0$ is defined after eq.(\ref{expo1}) in terms of the
momentum.
The theory can be quantized by imposing  equal time canonical commutation
relations
\beq
\label{canqu1}
[\dot X^\mu(\sigma,\tau),X^\nu(\sigma',\tau)]=-i\delta(\sigma-\sigma')
\eta^{\mu\nu},
\eeq
\beq
\label{canqu2}
[X^\mu(\sigma,\tau),X^\nu(\sigma',\tau)]=
[\dot X^\mu(\sigma,\tau),\dot X^\nu(\sigma',\tau)]=0~,
\eeq
which require the following commutation
relations on the oscillators and the zero modes
\beq
\label{oscqua1}
[\alpha^\mu_m,\alpha^\nu_n]=[\widetilde\alpha^\mu_m,\widetilde\alpha^\nu_n]=
m\delta_{m+n,0}\eta^{\mu\nu}~~~~~;~~~~~
[{\hat q}^\mu,{\hat p}^\nu]=i\eta^{\mu\nu}~,
\eeq
\beq
\label{oscqua2}
[\alpha^\mu_m,\widetilde\alpha^\nu_n]=[\hat q^\mu,\hat q^\nu]=
[\hat p^\mu,\hat p^\nu]=0~.
\eeq
In the quantum theory the Virasoro generators given in eqs.(\ref{vgenc}) and
(\ref{vgeno}) are defined by normal ordering the oscillators. But the only
operators for which this normal ordering matters
are  $L_0$ and ${\widetilde{L}}_0$, because they are the only ones
containing
products of non-commuting oscillators. We get therefore:
\beq
\label {L0c}
L_0 = \frac{\alpha '}{4}  {\hat{p}}^2 +
\sum_{n=1}^{\infty}  \alpha_{-n} \cdot \alpha_n
~~~~~~~~~~~
;
~~~~~~~~~~~
\widetilde L_0 = \frac{\alpha '}{4} {\hat{ p}}^2 +
\sum_{n=1}^{\infty}  \widetilde\alpha_{-n} \cdot \widetilde \alpha_n~~,
\eeq
for closed strings, and
\beq
L_0 = \alpha ' {\hat p}^2 +\sum_{n=1}^{\infty} \alpha_{-n} \cdot
\alpha_n~~,
\label{L0on}
\eeq
for open strings.
The commutation relations for the $L_n$ operators
give rise to the Virasoro algebra with
central extention
\beq
\label{VIR3}
[ L_m, L_n ]  =  (m-n) L_{m+n} + \frac{d}{12} m(m^2 -1) \delta_{m+n,o}~,
\eeq
The central
extension of Virasoro algebra is a consequence of the fact that we have
defined $L_0$ with the normal ordering. In the case of a closed string we
also have the
operators ${\widetilde{L}}_m$ that commute with all $L_m$ operators and
satisfy the
same Virasoro algebra as in eq.(\ref{VIR3}).

In the quantum theory the oscillators $\alpha_n$ and ${\widetilde\alpha}_n$
become creation and annihilation operators acting on a Fock space.
The vacuum state  $|0\rangle_\alpha |0\rangle_{\widetilde\alpha}|p\rangle$
with momentum $p$ is defined by the conditions
\[
\alpha^{\mu}_{n}|0\rangle_\alpha |0\rangle_{\widetilde\alpha}|p \rangle =
\tilde\alpha^{\mu}_{n} |0\rangle_\alpha |0\rangle_{\widetilde\alpha}
|p \rangle =0~~\forall n>0 ~~,~~
\]
\beq
\label{vacdef}
\hat p^\mu
|0\rangle_\alpha |0\rangle_{\widetilde\alpha}|p\rangle
=p^\mu |0\rangle_\alpha |0\rangle_{\widetilde\alpha}|p\rangle~,
\eeq
Because of the Lorentz metric
the Fock space defined by the commutation relations in eqs.(\ref {oscqua1})
and (\ref {oscqua2}) contains states with negative norm. The physical states
in the closed string case are characterized by the following conditions:
\beq
\label {phcond}
\left\{ \begin{array}{l}
L_m|\psi_{\rm phys}\rangle = {\widetilde{L}}_m|\psi_{\rm phys}\rangle = 0
~~~~~~m > 0\\
(L_0-1)|\psi_{\rm phys}\rangle = ({\widetilde{L}}_0-1)|\psi_{\rm
phys}\rangle =0~~,
\end{array}
\right.
\eeq
where the intercept $-1$ appearing in the second equations is a
consequence of the normal ordering of $L_0$ and $\widetilde L_0$.
In the open string we have to impose only
one set of the previous conditions.

From the lowest eqs. in  (\ref {phcond}) and from eqs.(\ref {L0c}) and
(\ref {L0on}) we can read the expression for the mass operator. For an open
string one gets
\beq
\label{massao}
 M^2=\frac{1}{\alpha'}\left(\sum_{n=1}^\infty
\alpha_{-n} \cdot \alpha_n-1\right) ,
\eeq
while for a closed string one gets
\beq
\label{massac}
M^2=\frac{2}{\alpha'}\left[\sum_{n=1}^\infty\left(
\alpha_{-n} \cdot \alpha_n+\widetilde\alpha_{-n} \cdot
\widetilde\alpha_n\right)-2
\right] ,
\eeq
together with the level matching condition:
\beq
( {\widetilde{L}}_0 - L_0 ) | \psi_{\rm phys} \rangle  =0.
\label{levmat}
\eeq
The action of superstring in the superconformal gauge is
\beq
\label {sstac}
S=-\frac{T}{2}\int_M d\tau d\sigma \left(\eta^{\alpha\beta} \partial_\alpha
X^\mu\partial_\beta X_\mu-i{\bar
\psi}^\mu\rho^\alpha\partial_\alpha\psi_{\mu}
\right),
\eeq
where $\psi$ is a world sheet Majorana spinor and the matrices
\beq
\label{romat}
\rho^0=\left( \begin{array}{cc} 0  & -i \\
                         i &  0 \end{array} \right)
~~~~~~~~
\rho^1=\left( \begin{array}{cc} 0  & i \\
                         i &  0 \end{array} \right),
\eeq
provide a representation of the Clifford algebra in two dimensions.
The previous action is invariant under the  following supersymmetry
transformations
\beq
\label{susyt}
\delta X^\mu=\bar\varepsilon\psi^\mu~~~
\delta \psi^\mu=-i\rho^\alpha\partial_\alpha X^\mu\varepsilon~~,
\eeq
where $\varepsilon$ is a constant Majorana spinor.
The N{\"{o}}ther current corresponding to the previous invariance is the
supercurrent
\beq
\label{scur}
J_\alpha=\frac{1}{2}\rho^\beta\rho_\alpha\psi^\mu\partial_\beta X_\mu.
\eeq
It is useful to write
the equations of motion for the fermionic degrees of freedom
in the light cone coordinates
\beq
\label{ssteq}
\partial_+\psi^\mu_-=0~~~~~;~~~~~\partial_-\psi^\mu_+=0,
\eeq
where
\beq
\label{psipm}
\psi^\mu_\pm=\frac{1\mp\rho^3}{2}\psi^\mu~~~~{\rm with}
~~\rho^3\equiv\rho^0\rho^1.
\eeq
The boundary conditions are
\beq
\label{sstbc}
\int d \tau \left(
\psi_+\delta\psi_+-\psi_-\delta\psi_-\right)|_{\sigma=0}^{\sigma=\pi}=0.
\eeq
As before, also these boundary conditions can be fulfilled in two different
ways. In the case of an open string eqs.(\ref{sstbc})
are satisfied if we require
\beq
\label{fbcu1}
\left\{
\begin{array}{l}
\psi_-(0,\tau)=\eta_1{\psi_+}(0,\tau)\\
\psi_-(\pi,\tau)=\eta_2{\psi_+}(\pi,\tau)
\end{array}
\right. ,
\eeq
where $\eta_1$ and $\eta_2$ can take the values $\pm 1$.
In particular if $\eta_1 =\eta_2$
we get what is called the Ramond (R) sector of the open string,
while if $\eta_1 =- \eta_2$ we get  the
Neveu-Schwarz (NS) sector.
\noindent
In the case of a closed string the fermionic coordinates $\psi_{\pm}$ are
independent from each other and they can be either periodic or
anti-periodic.
This amounts to impose the following  conditions:
\beq
\label {ucbc}
\psi^\mu_-(0,\tau)=\eta_3\psi^\mu_-(\pi,\tau)~~~~~
{\psi^\mu_+}(0,\tau)=\eta_4{\psi^\mu_+}(\pi,\tau),
\eeq
that satisfy the boundary conditions in eq.(\ref{sstbc}).
In this case we  have four different sectors
according to the two values that $\eta_3$ and $\eta_4$ take
\beq
\left\{
\begin{array}{l}
\eta_3=\eta_4=1\Rightarrow {\rm (R-R)}\\
\eta_3=\eta_4=-1\Rightarrow {\rm (NS-NS)}\\
\eta_3=-\eta_4=1\Rightarrow {\rm (R-NS)}\\
\eta_3=-\eta_4=-1\Rightarrow {\rm(NS-R)}\\
\end{array}\right. .
\label {rns2}
\eeq
The general solution of eq.(\ref{ssteq}) satisfying the boundary
conditions in eqs.(\ref{fbcu1}) is given by
\beq
\label{modpsil}
\psi^\mu_\mp \sim \sum_{t} \psi^\mu_t
e^{- it(\tau\mp\sigma)}~~~~~~~{\rm where}~~~~~
\left\{ \begin{array}{l}
t\in Z+\frac{1}{2}\rightarrow {\rm NS \,\,\,sector}\\
t\in Z\rightarrow {\rm R \,\,\, sector}
\end{array}
\right. ,
\eeq
while the ones satisfying the boundary conditions in eq.(\ref{ucbc}) are
given
by
\beq
\label{modpsil2}
\psi^\mu_- \sim \sum_{t} \psi^\mu_t
e^{-2it(\tau-\sigma)}~~~~~~~
{\rm where}~~~~~
\left\{ 
\begin{array}{l}
t\in Z+\frac{1}{2}\rightarrow {\rm NS \,\,\, sector}\\
t\in Z\rightarrow {\rm R \,\,\,sector}
\end{array}
\right. ,
\eeq
\beq
\label{modpsir}
\psi^\mu_+ \sim \sum_{t} \widetilde\psi^\mu_t 
e^{-2it(\tau +\sigma)}~~~~~~~
{\rm where}~~~~~
\left\{ 
\begin{array}{l}
t\in Z+\frac{1}{2}\rightarrow {\widetilde{\rm {NS}} \,\,\,{\rm sector}}\\
t\in Z\rightarrow {\widetilde{\rm {R}} \,\,\,{\rm sector}}
\end{array}
\right. .
\eeq
The energy-momentum tensor, in the light cone coordinates, 
has two non zero components
\beq
\label{sliccoem}
T_{++}=\partial_+ X \cdot \partial_+ X
+\frac{i}{2}\psi_+ \cdot \partial_+\psi_{+}
~~~~;~~~~
T_{--}=\partial_- X \cdot \partial_- X
+\frac{i}{2}\psi_- \cdot \partial_-\psi_{-}, 
\eeq
while the supercurrent  
defined in eq. (\ref{scur}) reduces to 
\beq
\label{scurpm}
J_-=\psi_- \cdot \partial_-X ~~~~;~~~~J_+=\psi_+ \cdot \partial_+X .
\eeq
From the energy-momentum tensor we can get the Virasoro generators using again
the mode expansion in eq.(\ref{liccoem}) and one gets 
\beq
\label{svgeno}
L_{n}=\frac{1}{2}\sum_{m\in Z}\alpha_{-m}\cdot \alpha_{n+m}
+
~\frac{1}{2}\sum_t\left(\frac{n}{2}+t\right)\psi_{-t} \cdot \psi_{t+n},
\eeq
for an open string and 
\[
L_{n}=\frac{1}{2}\sum_{m\in Z}\alpha_{-m} \cdot \alpha_{n+m}+
\frac{1}{2}\sum_t\left(\frac{n}{2}+t\right)\psi_{-t} \cdot \psi_{t+n},
\]
\beq
\label{svgenc}
\widetilde L_{n}=\frac{1}{2}\sum_{m\in Z}\widetilde\alpha_{-m} \cdot \widetilde\alpha_{n+m}
+ \frac{1}{2}\sum_t\left(\frac{n}{2}+t\right)\widetilde\psi_{-t} \cdot
\widetilde\psi_{t+n},
\eeq
for a closed string. The index $t$ used in the previous expressions and the
index $v$ that will be used later on refer both to 
the NS sector where $t\in Z+\frac{1}{2}$ and to the R sector where
$t\in Z$.

The Fourier components of the supercurrent that we
denote with $G_t$ and $\widetilde G_t$ are given by
the following expressions in terms of the oscillators
\beq
\label{sucurfu}
G_t=\sum_{n=-\infty}^\infty\alpha_{-n} \cdot \psi_{t+n}~~~~;~~~~
\widetilde G_t=\sum_{n=-\infty}^\infty\widetilde\alpha_{-n} \cdot
\widetilde\psi_{t+n}.
\eeq
The superstring can be quantized by imposing the canonical commutation
relations in eqs.(\ref{canqu1}) and (\ref{canqu2}) for the bosonic
coordinates and the following canonical anticommutation relations for the
fermionic ones
\beq
\label{ferqua}
\{\psi^\mu_A(\sigma,\tau),\psi^\mu_B(\sigma'\tau)\}=\pi\delta(\sigma-\sigma'
)
\eta^{\mu\nu}\delta_{AB}.
\eeq
In terms of the oscillators, together with eqs.(\ref{oscqua1}) and
(\ref{oscqua2}) we have
\beq
\label{ferquosc}
\{\psi^\mu_t,\psi^\nu_v\}=
\eta^{\mu\nu}\delta_{v+t,0}.
\eeq
Also in the supersymmetric case the quantum Virasoro generators are defined
with a normal ordered product of the oscillators, and again the normal
ordering affects only the $L_0$ operator that becomes
\beq
L_0 = \alpha ' \hat p^2 +\sum_{n=1}^{\infty} \alpha_{-n} \cdot \alpha_n 
+\sum_{t >0} t \psi_{-t} \cdot \psi_t~.
\label{sL0on}
\eeq
in the case of an open string, while in the case of a closed string the 
operators $L_0$ and ${\tilde{L}}_0$ are given by
\beq
L_0 = \frac{\alpha '}{4} \hat p^2 +\sum_{n=1}^{\infty} \alpha_{-n} \cdot 
\alpha_n +\sum_{t >0} t \psi_{-t} \cdot \psi_t
\label{L0v}
\eeq
and
\beq
{\tilde{L}}_0 = \frac{\alpha '}{4} \hat p^2 +\sum_{n=1}^{\infty} 
{\tilde{\alpha}}_{-n} \cdot 
{\tilde{\alpha}}_n +\sum_{t >0} t {\tilde{\psi}}_{-t} \cdot {\tilde{\psi}}_t
\label{L0v1}
\eeq
 The (anti)commutation relations for the operators  given
in eqs. (\ref{svgeno}), for $n\neq 0$ and
 (\ref{sL0on}) for $n=0$ and the operators
(\ref{sucurfu}) give rise to the super Virasoro algebra with
central extention
\beq
\label{b26}
\begin{array}{lll}
\,[ L_m, L_n ] & = & (m-n) L_{m+n} + \frac{d}{8} m(m^2 -1) \delta_{m+n,o} \\ 
\, [ L_m, G_r ] & = & (\frac{1}{2} m - r) G_{r+m}  \\ 
\, \{ G_r, G_s \} & = & 2L_{r+s} + 
 \frac{d}{2} (r^2 - \frac{1}{4}) \delta_{r+s,o} 
\end{array}\,\,\,\,\,\,\,\,\,\,\,\,\,\,(NS)
\eeq
for the NS sector and
\beq
\label{b27}
\begin{array}{lll}
\,[ L_m, L_n ] & = & (m-n) L_{m+n} + \frac{d}{8} m^3 \delta_{m+n,o}  \\ 
\, [ L_m, G_n ] & = & (\frac{1}{2} m - n) G_{n+m}  \\ 
\,\{ G_m, G_n \} & = & 2L_{m+n} + 
 \frac{d}{2} n^2 \delta_{m+n,o} 
\end{array}\,\,\,\,\,\,\,\,\,\,\,\,\,\,(R)
\eeq
in the R sector.  Notice that only the
c-number terms in the r.h.s. of the previous equations
are different in the two sectors. The algebra of the Ramond sector
can be brought into the same form as the one in the NS sector by a
redefinition
of $L_0 \rightarrow L_0 + d/16$. This observation will be used later on to
determine the dimension of the spin field operator in the Ramond sector.

Also in  superstring the spectrum contains unphysical states with
negative
norm. The conditions which select the physical states are
\beq
\label {sphcond1}
\left\{ \begin{array}{l}
L_m|\psi_{\rm phys}\rangle =0 ~~~~~m > 0\\
(L_0-a_0 )|\psi_{\rm phys}\rangle =0\\
G_t|\psi_{\rm phys}\rangle =0 ~~~~~\forall t \geq 0\\
\end{array}
\right. ,
\eeq
where
\beq
\label {phcond2}
\left\{ \begin{array}{l}
a_0 =\frac{1}{2}~~~~~{\rm for~ the~NS~sector}\\
a_0 =0~~~~~{\rm for~ the~R~sector}
\end{array}
\right. .
\eeq
In the case of a closed string we should add to the previous conditions the
analogous ones involving the tilded sector.
\noindent
From the middle condition in eq. (\ref{sphcond1}) we can read the expression
of the mass operator, that in the open string case is equal to
\beq
\label{smassao}
M^2=\frac{1}{\alpha'}\left(\sum_{n=1}^{\infty}  \alpha_{-n} \cdot \alpha_n
+\sum_{t>0} t\psi_{-t} \cdot \psi_t-a_0 \right).
\eeq
In the case of a closed string we get instead
\beq
\label{smassacl}
M^2= \frac{1}{2} \left( M^2_+ +M^2_- \right)~~,
\eeq
where
\beq
\label{smassacl1}
M^2_-=\frac{4}{\alpha'}\left(\sum_{n=1}^{\infty}  \alpha_{-n} \cdot \alpha_n
+\sum_t t\psi_{-t} \cdot \psi_t-a_0 \right),
\eeq
\beq
\label{smassacl2}
M^2_+=\frac{4}{\alpha'}\left(
\sum_{n=1}^{\infty}  \widetilde\alpha_{-n} \cdot
\widetilde\alpha_n
+\sum_t t\widetilde\psi_{-t} \cdot \widetilde\psi_t-{\tilde{a}}_0 \right),
\eeq
and the values of $a_0$ and ${\tilde{a}}_0$ are given in eq.(\ref{phcond2}) 
depending if
we are in the NS or in R sector. In the closed string case we should also
add
the level matching condition
\beq
(L_0 - {\widetilde{L}}_0 -a_0  +{\tilde{a}}_0 ) | \psi_{\rm phys} \rangle  =0
\label{levmat4}.
\eeq

\section{Conformal Field Theory Formulation}
\label{sec:third}

As mentioned in Sect.~\ref{sec:secondsection}, string theories in the conformal 
gauge are two-dimensional conformal field theories.
Thus, instead of the operatorial analysis that we have discussed
until now, one can give an equivalent description by using the language of 
conformal  field theory in which one works with the OPE
rather then commutators or anticommutators and that contributes to simplify
many calculations. 
In the case of a closed string it is convenient to introduce the variables
$z$ and ${\bar{z}}$ that are related to the world sheet variables $\tau$ and
$\sigma$ through a conformal transformation:
\beq
z = e^{ 2i( \tau - \sigma)}~~~~~~~~~~;~~~~~~~~~~ {\bar{z}} =
e^{ 2 i( \tau + \sigma)}~,
\label{zbarz}
\eeq
In the case of an euclidean world sheet ($\tau \rightarrow -i \tau$) 
$z$ and ${\bar{z}}$ are complex conjugate of each other. In terms of them we 
can write the bosonic coordinate $X^{\mu}$ as follows:
\beq
X^{\mu} (z, {\bar{z}})  = \frac{1}{2} \left[X^{\mu} (z) +
{\widetilde{X}}^{\mu}
( {\bar{z}}) \right]
\label{Xzbarz}
\eeq
where
\beq
X^{\mu} (z) = {\hat{q}}^{\mu} - i \sqrt{2 \alpha'} \log z \alpha_{0}^{\mu} + i
\sqrt{2 \alpha'} \sum_{n \neq 0} \frac{\alpha_{n}^{\mu}}{n} z^{-n}
\label{Xmuz}
\eeq
and
\beq
{\widetilde{X}}^{\mu} ({\bar{z}}) = {\hat{q}}^{\mu} - i \sqrt{2 \alpha'} \log
{\bar{z}}
{\widetilde{\alpha}}_{0}^{\mu} + i
\sqrt{2 \alpha'} \sum_{n \neq 0} \frac{{\widetilde{\alpha}}_{n}^{\mu}}{n}
{\bar{z}}^{-n}
\label{Xmuzbar}
\eeq
with
$\alpha^{\mu}_{0} = {\widetilde{\alpha}}^{\mu}_{0} =
\sqrt{\alpha' /2}\,\, {\hat{p}}^{\mu}$.
In the case of an open string theory one can introduce the
variables:
\beq
z = e^{i (\tau-\sigma)}~~~;~~~ \bar z=e^{i (\tau+\sigma)}
\label{zopen}
\eeq
and the string coordinate can be written as
\beq
\label{opecor}
X^{\mu} (z, {\bar{z}})  = \frac{1}{2} \left[X^{\mu} (z) +
{{X}}^{\mu}
( {\bar{z}}) \right]~~,
\eeq
where  $X^\mu$ is given in eq. (\ref{Xmuz})
and $\sqrt{2 \alpha'}{\hat{p}}^{\mu} = \alpha_{0}^{\mu}.$
In superstring theory we must also introduce a conformal field with
conformal dimension equal to $1/2$ corresponding to the fermionic coordinate.
In the closed string case we have two independent fields for the holomorphic 
and anti-holomorphic sectors which are obtained from eqs. (\ref{modpsil2}) 
and (\ref{modpsir}) through the Wick rotation $\tau\rightarrow -i\tau$ and
the conformal transformation $(\tau,\sigma)\rightarrow(z,\bar z)$
\beq
\label{fcoo}
\Psi^{\mu} (z) \sim
\sum_{t}\psi_{t} z^{-t- 1/2}~~~;~~~
\widetilde\Psi^{\mu} (\bar z) \sim
\sum_{t} \widetilde\psi_{t} \bar z^{-t- 1/2}
\eeq
In the open string case, starting from eq.(\ref{modpsil}) and applying
the same operations we get again eqs.(\ref{fcoo}), but this time with the
same oscillators.

In what follows we will explicitly consider only 
the holomorphic sector for the closed
string. Analogous considerations hold for the antiholomorphic sector.
In the case of an open string it is sufficient to consider the string 
coordinate at the string endpoint $\sigma =0$. In both cases  
it is convenient to introduce  a bosonic dimensionless 
variable:
\beq
\label{bcoo}
x^{\mu} (z) \equiv X^{\mu} (z)/(\sqrt{2 \alpha'}) = {\tilde{q}}^{\mu} -
i\alpha_{0}^{\mu}
\log z + i \sum_{n \neq 0} \frac{\alpha_n}{n} z^{-n}~~,
\eeq
where ${\tilde{q}} = {\hat{q}}/\sqrt{2 \alpha'}$ and a fermionic one:
\beq
\psi^{\mu} (z) = -i \sum_{t} \psi_{t} z^{-t -1/2}
\label{psi76}
\eeq 
The theory can be quantized by imposing the following OPEs
\beq
 x^{\mu} (z)  x^{\nu} (w)   = - \eta^{\mu \nu} \log (z -w) 
+...;~~~  
\psi^{\mu} (z) \psi^{\nu} (w)  = - \frac{\eta^{\mu
\nu}}{z -w}+...~~,
\label{contra}
\eeq
where the dots denote finite terms for $z\rightarrow w$.
Notice that these OPEs coincide with the $2$-points Green's functions except 
for the Ramond case where the Green's function is equal to:
\beq
\label{Ram87}
< \psi^{\mu} (z) \psi^{\nu} (w) > = - \frac{\eta^{\mu \nu}}{z -w}
\frac{1}{2}
\left[\sqrt{\frac{z}{w}} + \sqrt{\frac{w}{z}} \right]~~.
\eeq
Since, however, its singular behaviour when $z \rightarrow w$ is the same as
in
eq.(\ref{contra}), we use the contractions in eqs.(\ref{contra}) for both
the
NS and the R sector.
In terms of the previous conformal fields we can define the generators of
superconformal transformations:
\beq
G(z) = -\frac{1}{2}\psi \cdot \partial x ~~~~;~~~~
T(z) = T^x(z)+T^\psi(z)=
- \frac{1}{2} (\partial x )^2 - \frac{1}{2} \partial \psi \cdot \psi~~.
\label{gl}
\eeq
Their mode expansion is given by:
\beq
L_n = \frac{1}{2 \pi i} \oint dz z^{n+1} T(z) \hspace{1cm}; \hspace{1cm}
G_t = \frac{1}{2 \pi i} \oint dz z^{t + 1/2    }  G(z)~~.
\label{moex}
\eeq
The conformal fields in eq.(\ref{gl}) satisfy the following OPEs:
\beq
T(z) T(w) = \frac{ \frac{d}{dw} T(w)}{z-w} + 2 \frac{T(w)}{(z-w)^2} +
\frac{c/2}{(z-w)^4}+...~~,
\label{ll}
\eeq
\beq
T(z)  G(w) = \frac{\partial/\partial w G(w)}{z-w} + \frac{3}{2} 
\frac{G(w)}{(z-w)^2}+...
\label{lg}
\eeq
\beq
\label{gg}
 G(z)  G(w)   = \frac{2 T(z)}{z-w} + \frac{d}{(z-w)^3}+...~~.
\eeq
Using eqs.(\ref{moex}) it is easy to see that the previous OPEs imply the
super Virasoro algebra in eq.(\ref{b26}) for both the NS and the R sector. 
But then the superconformal algebra that we get in the R sector 
differs from the one given in eq.(\ref{b27}). However, 
as we have noticed in the previous section,
eq.(\ref{b27}) can be reduced to eq. (\ref{b26})  
by translating $L_0$ in the R sector in eq.({\ref{b27}) by a constant:
\beq
L_0 \rightarrow L_{0}^{conf} \equiv  L_0 +\frac{d}{16}=
\sum_{n=1}^{\infty} \left( \alpha_{-n} \cdot \alpha_n + n \psi_{-n}
\cdot \psi_n  \right) + \alpha' p^2 + \frac{d}{16}
\label{lonew4}
\eeq
Therefore in the R sector we have two $L_0$ operators that are related
by eq.(\ref{lonew4}). $L_0$ determines the spectrum of 
superstring through eq.(\ref{sphcond1}),
while $L_{0}^{conf}$ that satisfies the algebra in eq.(\ref{b26}) encodes 
the correct conformal properties of the R sector.

In conformal field theory one introduces the concept of conformal or primary 
field 
$ \Phi (z)$ with dimension $h$ as the object that satisfies the following OPE
with the energy-momentum tensor:
\beq
\label{confi}
T(z) \Phi (w)= \frac{\partial_w \Phi(w)}{z-w}+
h \frac{ \Phi (w)}{(z-w)^2} +...~~.
\eeq
From it one can compute the corresponding highest weight state 
$| \Phi \rangle$ by means of the following limiting procedure 
\beq
\label{vertice}
|\Phi \rangle=
\lim_{z\rightarrow 0} \Phi (z)|0\rangle~~,~~~~
\langle\Phi|=\lim_{z\rightarrow 0}\langle 0| \Phi^\dagger (z) \sim
\lim_{z\rightarrow \infty}\langle 0|( z^2)^h \Phi (z)
\eeq
The hermitian conjugate field ${\Phi}^\dagger$
in the previous expression has been defined as the field transformed under 
the conformal transformation $z\rightarrow 1/z$ apart from possibly a phase 
factor.
Using the previous definition and the expression for $L_0$ given in 
eq.(\ref{moex}), it is easy to see that, if the conformal field has
conformal dimension $h$, then the corresponding state is an eigenstate of $L_0$
with eigenvalue $h$ and it is annihilated by all the Virasoro operators with
$n>0$, namely 
\beq
L_0|\Phi\rangle=h|\Phi\rangle ~~~~~;~~~~L_n |\Phi\rangle =0~~~~.
\label{primary}
\eeq 
In the bosonic case the physical conditions on the states given in 
eq.(\ref{phcond}) imply that the vertex operators of the bosonic string theory 
${\cal V}_\alpha (z)$ are conformal fields with conformal dimension equal to 
$1$.
This insures that the quantity $dz {\cal V}_{\alpha} (z)$ is invariant
under conformal transformations.
In the supersymmetric case the physical conditions in eq.(\ref{sphcond1})
imply $L_0|\psi_{phys}\rangle=1/2|\psi_{phys}\rangle$ in the NS-sector and 
$L_0|\psi_{phys}\rangle=0$ in the R-sector. Therefore the 
corresponding vertex operators are not conformal fields with conformal dimension
equal to $1$ as in the bosonic string. On the other hand in the case
of the R sector we have seen that,
in order to determine the correct dimension of a vertex operator, we should
use the operator $L_0^{\rm conf}$ that, for $d=10$, acts on 
the spinorial ground state $| A \rangle$ as follows
\beq
L_0^{conf} | A \rangle  = \frac{5}{8} |A \rangle ~,
\label{condim3}
\eeq
This does not help, however, to get a vertex operator that has dimension equal
to $1$. In the case of superstring we will see that, 
in order to get the correct physical vertex operators in both  NS and R sectors,
one must  add the contribution of the superghost degrees of freedom that
will be discussed later on. The R vacuum state $| A \rangle $ in 
eq.(\ref{condim3}) can be written in terms of the NS
vacuum by introducing the spin field operator $S^{A} (z)$  satisfying the eq.:
\beq
\lim_{z \rightarrow 0} S^{A} (z) |0\rangle  = |A\rangle~~,
\label{spfi86}
\eeq
where $|0\rangle $ is the NS vacuum.
Thus from eq. (\ref{condim3}) we see that the spin field $S^A(z)$, which maps 
the NS vacuum into the R one, must have confromal dimension $5/8.$ 

One can show that the spin field $S^A(z)$ satisfies the following 
OPEs~\cite{FMS}:
\[
\psi^\mu(z)S_A(w)=(z-w)^{-1/2}(\Gamma^\mu)_{AB}S^B(w)+...
\]
\beq
\label{spis}
S_A(w)S_B(w)=(z-w)^{-3/2}(\Gamma)_{AB}\psi^\mu+...~~.
\eeq
Until now we have completely disregarded the analysis of the ghost and 
superghost degrees of freedom that, however, must be included in a 
correct Lorentz covariant quantization of string theory.
They arise from the exponentiation of the Faddev-Popov determinant
that is obtained when the string is quantized through 
the path-integral quantization.
In particular in the bosonic case, choosing the conformal gauge, one gets
the following ghost action~\cite{gsw}
\beq
\label{Sbc} 
S_{\rm ghosts} \sim \int d^2 z[b\bar\partial c+{\rm c.c.}]~~,
\eeq
where $b$ and $c$ are fermionic fields with conformal dimension equal 
respectively to $2$ and $-1$. 
The ghost system of the bosonic string is a particular case of the fermionic
$bc$ system described in the Appendix corresponding to a  screening
charge ${\cal Q}=-3$ and a central charge of the Virasoro operator $c=-26.$

With the introduction of  ghosts the string action  in the 
conformal gauge becomes invariant under the BRST transformations and
the physical states are  characterized by the fact that they are annihilated 
by the BRST charge that in the bosonic case is given by
\beq
\label{brchar}
Q\equiv
\oint\frac{dz}{2\pi i}J_{BRST}(z)\equiv
\oint\frac{dz}{2\pi i}c(z)\left[T^x(z)+\frac{1}{2}T^{bc}(z)
\right]~~,
\eeq
where $T^x(z)=-1/2(\partial x)^2,$ 
and $T^{bc}$ is given in eq.(\ref{2.6}) for 
$\lambda =2$. 
It can be shown that $Q$ is nilpotent  if the space-time dimension $d=26$.
The physical states are annihilated by the BRST charge 
\beq
\label{phisst}
Q|\psi_{\rm phys}\rangle=0
\eeq 
This implies that the vertex operators corresponding to the physical states
must satisfy the condition
\beq
\label{verst}
 [Q,{\cal W}(w)]_\eta=\oint_{w}
\frac{dz}{2\pi i}J_{BRST}(z)\, {\cal W}(w)=0~~,
\eeq  
where $[,]_\eta$ means  commutator ($\eta =-1$) [anticommutator ($\eta =1$)] 
when the vertex operator is a bosonic [fermionic] quantity. 
By using the OPE it can be shown that in the bosonic string the most general 
BRST invariant vertex operator has  the following form
\beq
\label{binvert}
{\cal W}(z)=
c(z){\cal V}^x_\alpha(z)
\eeq
where ${\cal V}^x_\alpha$ is a conformal field with dimension equal to 1
that depends only on the string coordinate $x^{\mu}$.

In the supersymmetric case one must add to the ghost action 
in eq.(\ref{Sbc}) the superghost one: 
\beq
\label{sSbc}
S_{sghost} \sim \int d^2 z\left(\beta\bar\partial\gamma+c.c\right)~~,
\eeq
where  $\beta$ and $\gamma$  are bosonic fields 
with conformal dimensions equal respectively to $3/2$ and
$-1/2$. This is a particular case of the bosonic $bc$ system described in the 
Appendix  with $\epsilon=-1$, $\lambda=3/2$ and  ${\cal Q}=2$, corresponding 
to a central charge of the Virasoro algebra $c=11$.
The BRST charge can be conveniently defined by using a world-sheet
superfield formulation where one introduces 
\beq
\label{sufil}
Z=(z,\theta)~~~;~~~
\hat X(Z)=x+\theta\psi~~;~~ D=\partial_\theta+\theta\partial_z
\eeq
and defines the BRST-supercharge as
\beq
\label{sbrchar}
Q=\oint\frac{dz d\theta}{2\pi i}C(Z)\left[\hat T^m(Z)+\frac{1}{2}\hat T^g(Z)
\right]
\eeq
where 
\beq
\label{sufilem}
\hat T^m(Z)=-\frac{1}{2} D\hat X\partial \hat X
=G(z)+\theta T(z)~~,
\eeq
\beq
\label{sufilcb}
C(Z)=c(z)+\theta\gamma(z)~~~;~~~ B(Z)=\beta(z)+\theta b(z)
\eeq
with $T(z)$ and $G(z)$ defined in eq.(\ref{gl}) and 
\[
\hat T^g(Z) \equiv G^g (z) + \theta T^{g} (z) =
-C\partial B +\frac{1}{2}DCDB-\frac{3}{2}\partial C B =
\]
\beq
\label{sufilem2}
=-c\partial\beta+\frac{1}{2}\gamma\beta-\frac{3}{2}\partial c\beta
+\theta (T^{bc}+T^{\beta\gamma})~~.
\eeq
Performing the Grassmann integration over $\theta$ one gets
\beq
\label{carS}
Q \equiv \oint dz J_{BRST} (z) = Q_0+Q_1+Q_2~~,
\eeq
where
\beq
\label{cars2}
Q_0= \oint\frac{dz}{2\pi i}c(z)\left[T(z) +T^{\beta\gamma}(z)+
\partial c(z)b(z)
\right]
\eeq
and
\beq
\label{cars3}
Q_1=\frac{1}{2}\oint\frac{dz}{2\pi i} \gamma(z)\psi(z)\cdot\partial X(z)~~~;
~~~ Q_2=-\frac{1}{4}\oint\frac{dz}{2\pi i}\gamma^2(z)b(z)
\eeq
A vertex operator corresponding to a physical state must be BRST invariant,
i.e.
\beq
\label{verstsu}
 [Q,{\cal W} (Z)]_\eta=0 
\eeq
Before the introduction of  ghosts and superghosts, the vertex operators
for the NS sector in the superfield formalism can be written as
\beq
\label{vert1}
{\cal V}(Z)={\cal V}_0(z)+\theta {\cal V}_1(z)
\eeq
where ${\cal V}_0$ and ${\cal V}_1$  are two conformal fields with 
dimension $1/2$  
and $1$ respectively. For example in the massless NS sector the two fields
are given by
\beq
\label{ver2}
{\cal V}_0(z)=\epsilon\cdot\psi(z) e^{ik\cdot X(z)}~~;~~
{\cal V}_1(z)=(\epsilon\cdot\partial X(z) +ik\cdot\psi \epsilon\cdot\psi)
e^{ik\cdot X(z)}~~.
\eeq
But they are not BRST invariant. In order to construct a BRST invariant version
of the vertex ${\cal V}_0(z)$ we must add the contribution of the ghosts and
superghosts. This can be easily done and one gets
\beq
\label{vert3}
{\cal W}_{-1}(z)=c(z) e^{-\varphi(z)}{\cal V}_0(z)
\eeq
In the case of the massless vertex in eq.(\ref{ver2}) the vertex in  
eq.(\ref{vert3}) is BRST invariant  if $k^2= \epsilon\cdot k=0.$ 
We can proceed in an analogous way in the R sector and obtain the following
BRST invariant vertex operator for the massless fermionic state of  open
superstring~\cite{FMS}:
\beq
\label{vert7}
{\cal W}_{-1/2}(z)=u_A(k) c(z) S^A(z)e^{-\frac{1}{2}\varphi(z)} e^{ik\cdot X
(z)}
\eeq
It is BRST invariant if $k^2 =0$ and $u_A 
\left( \Gamma^{\mu} \right)^{A}_{\,\,B} k_{\mu} =0$. Both  vertices in 
eqs.(\ref{vert3}) and (\ref{vert7}) have conformal dimension equal to zero
as in the case of the bosonic string (see eq.(\ref{binvert})). 

In superstring, however, unlike the bosonic string, for each physical state
we can construct an infinite tower of equivalent physical vertex operators 
all (anti)commuting
with the BRST charge and characterized  according to their superghost picture
$P$ that is equal to the total ghost
number of the scalar field $\varphi$ and of the $\eta\xi$ system that appear 
in the "bosonization" of the $\beta\gamma$ system 
(see the Appendix for details):
\beq
P = \oint \frac{dz}{2 \pi i} \left(- \partial \varphi + \xi \eta \right)
\label{pict}
\eeq
Notice that the vertex in eq.(\ref{vert3}) is in the picture $-1$, while the 
one in eq.(\ref{vert7}) is in the picture $-1/2$.
Vertex operators in different pictures are related through the picture 
changing procedure that we are now going to describe. Starting from
a BRST invariant vertex ${\cal{W}}_{t}$ in the picture $t$ 
(characterized by a value of $P$ equal to $t$), where $t$ is 
integer (half-integer) in the NS (R) sector,
one can construct another BRST invariant vertex operator ${\cal W}_{t+1}$ in 
the picture  $t+1$ through the following operation~\cite{FMS}
\beq
\label{vert4}
{\cal W}_{t+1}(w)=
[Q,2\xi(w){\cal W}_{t}(w)]_{\eta} =\oint_{w}
\frac{dz}{2\pi i}J_{BRST}(z)~2\xi(w){\cal W}_{t}(w)~~.
\eeq 
Using the Jacobi identity and the fact that $Q^2 =0$ one can easily show that
the vertex ${\cal W}_{t+1} (w)$ is BRST invariant:
\beq
[Q , {\cal W}_{t+1} ]_{\eta} =0
\label{brsinv}
\eeq
On the other hand the vertex ${\cal W}_{t+1} (w) $ obtained through the
construction in eq.(\ref{vert4}) is not BRST trivial because the corresponding
state contains the zero mode $\xi_0$ that is not contained in the Hilbert
space of the $\beta\gamma$-system (see eq.(\ref{bcboso})). In conclusion all 
the vertices constructed
through the procedure given in eq.(\ref{vert4}) are BRST invariant and non
trivial in the sense that all give a non-vanishing result when inserted for 
instance in a
tree-diagram correlator  provided that the total picture number is equal to
$-2$.
Using the picture changing procedure from the vertex operator in 
eq.(\ref{vert3})
we can construct the vertex operator in the $0$ superghost picture which is 
given by~\cite{PETTO}
\beq
\label{vert6}
{\cal W}_{0}(z)=c(z) {\cal V}_1(z)-\frac{1}{2}\gamma(z){\cal V}_0(z)~~.
\eeq
Analogously starting from the massless vertex in the R sector in 
eq.(\ref{vert7}) one can construct the corresponding vertex in an arbitrary 
superghost picture $t$.

In the closed string case the vertex operators 
are given by the product of two vertex operators of the open string.
Thus for the massless NS-NS sector in the superghost picture $(-1,-1)$ 
we have
\beq
\label{vert7b}
{\cal W}_{(-1,-1)}=\epsilon_{\mu\nu}{\cal V}^\mu_{-1}(k/2,z)
\widetilde{\cal V}^\nu_{-1}(k/2,\bar z)~~,
\eeq 
where ${\cal V}^\mu_{-1}(k/2,z)=c(z)\psi^\mu(z)e^{-\varphi(z)}
e^{i\frac{k}{2}\cdot X(z)}$
and  $\widetilde{\cal V}^\nu_{-1}$ is equal to an analogous expression
in terms of the tilded modes. This vertex is BRST invariant if $k^2=0$ 
and $\epsilon_{\mu\nu}k^\nu=k^\mu\epsilon_{\mu\nu}=0.$

In the R-R sector the vertex operator for  massless states 
in the  $(- \frac{1}{2},- \frac{1}{2})$ superghost picture is
\beq
\label{vert8}
{\cal W}_{(-1/2,-1/2)}= 
\frac{\left(C\Gamma^{\mu_1...\mu_{m+1}}\right)_{AB}
F_{\mu_1...\mu_{m+1}}}{2\sqrt{2} (m+1)!}{\cal V}^A_{-1/2}(k/2,z)
\widetilde{\cal V}^B_{-1/2}(k/2,\bar z)
\eeq
where
${\cal V}^A_{-1/2}(k/2,z)=c(z)S^A(z)e^{-\frac{1}{2}\varphi(z)}
e^{i\frac{k}{2}\cdot X(z)}$ and
\beq
\label{effe}
F_{\mu_1...\mu_{m+1}}=\frac{(-1)^{m+1}}{2^5}u_D(k)(\Gamma_{\mu_1...\mu_{m+1}}
C^{-1})^{DE}\widetilde u_E(k)~~.
\eeq
It is BRST invariant if $k^2=0$ and $F_{\mu_1...\mu_m}$ is a field 
strength satisfying both the Maxwell equation $(dF=0)$
and the Bianchi identity $(d\,^*F=0)$.
The two Weyl-Majorana spinors $u_A$ and $\widetilde u_B$ 
may have the same or opposite chirality.
In the first case one obtains  type IIB theory
while in the second case one obtains the type IIA theory. 
From eq.(\ref{effe})
one can see that the only field strengths which are allowed
are those for $(m+1)$ odd in 
IIB theory and those for $(m+1)$ even in  IIA theory.
Moreover, from eq.(\ref{effe}) one can show that the field strenghts with
values of $m$ related by Hodge duality are not independent and one can
restrict oneself to the values  $(m+1)=1,~3,~5$  in type IIB  and
$(m+1)= 2,4$ in type IIA string theory.

Since the physical state corresponding to the symmetric vertex given in 
eq.(\ref{vert8}) cannot be used to compute its coupling with a D$-$ brane
because the boundary state that we will construct in Sect.~\ref{sec:seventh}
is in an asymmetric picture, in the following we will  explicitly write the 
vertex operator of a physical R-R state in the asymmetric picture
$(-1/2,-3/2)$. It is given by~\cite{BILLO}:
\beq
\label{vert9}
{\cal W}_{(-1/2,-3/2)}=\sum_{M=0}^{\infty} \frac{a_M}{2{\sqrt 2}}
\left(C {\cal{A}}^{(m)} \Pi_M\right)_{AB}
{\cal V}^A_{-1/2+M}(k/2,z)\widetilde{\cal V}^B_{-3/2-M}(k/2,\bar z)
\eeq
where
\beq
\left(C {\cal{A}}^{(m)} \right)_{AB} = \frac{\left(C \Gamma^{\mu_1...\mu_m}
\right)}{m!} A_{\mu_1...\mu_m}~~,~~\Pi_q=\frac{1+(-1)^q\Gamma_{11}}{2}  
\label{aaa}
\eeq
 and
\beq
\label{vert10}
{\cal V}^A_{-1/2+M}(k/2,z)=\partial^{M-1}\eta(z)...\eta(z)
c(z) S^A(z)e^{\left(-\frac{1}{2}+M\right)\varphi(z)} e^{i\frac{k}{2}\cdot X(z)}
\eeq
\beq
\label{vert11}
\widetilde{\cal V}^B_{-3/2-M}(k/2,\bar z)=
\bar\partial^{M}\widetilde\xi(\bar z)...\bar\partial\widetilde\xi(\bar z)
\widetilde c(\bar z) \widetilde S^A(\bar z)e^{\left(-\frac{3}{2}-M\right)
\widetilde\varphi(\bar z)} e^{i\frac{k}{2}\cdot \widetilde X(\bar z)}
\eeq
It can be shown that the vertex operator in eq.(\ref{vert9}) is BRST invariant
if $k^2=0$ and the following two conditions are satisfied
\beq
a_M= \frac{(-1)^{M(M+1)}}{[M!(M-1)!...1]^2}~~~~,~~~~d {}^* A^{(m)} =0~~~.
\label{condi8}
\eeq
By acting with the picture changing operator on the vertex in eq.(\ref{vert9})
it can be shown that one obtains the vertex in the symmetric picture in
eq.(\ref{vert8}). In particular one can show that only the first 
term in the sum in eq.(\ref{vert9})  reproduces the symmetric 
vertex, while all the other terms  give BRST trivial contributions.

\section{T-Duality}
\label{sec:fourth}

The compactification of a dimension in string theory is characterised by the
appearance of new interesting phenomena with respect to those already
present in field theory.
In fact, in the case of a closed string, together with the Kaluza-Klein
(K-K)
excitations, a new kind of states called winding states appear in the
spectrum. It turns out that the bosonic closed string theory is invariant
under the exchange of the winding modes with the K-K modes according
to a transformation that is called T-duality. In the supersymmetric case,
instead, this transformation is in general not a symmetry anymore but brings
from a certain string theory to another string theory. For instance T-duality
along a certain direction acts interchanging the IIA with the IIB theory.
In the case of an open string, instead, this analysis
naturally leads to the existence of other objects
called {\dpbs}.

Let us discuss in some detail the compactification and the T-duality
invariance,
starting with the bosonic closed string.

The most general solution of the eqs. of motion for the bosonic closed
string
in eq.(\ref {steq}) can be written as:
\[
X^\mu(\tau, \sigma)= q^\mu+\sqrt{
2\alpha'}\left( \alpha_0^\mu+{\widetilde\alpha}_0^\mu\right) \tau-
\sqrt{2\alpha'}\left( \alpha_0^\mu-{\widetilde\alpha}_0^\mu\right)
\sigma +
\]
\beq
\label {gexpc}
+i\sqrt{\frac{\alpha'}{2}}
\sum_{n\neq 0}
\left(\frac {\alpha^{\mu}_{n}}{n}e^{-2in(\tau -\sigma)}+
\frac {\widetilde\alpha^{\mu}_{n}}{n}e^{-2in(\tau+\sigma)}\right)~,
\eeq
where the momentum of the string is given by
\beq
\label{gmom}
p^\mu=\frac{1}{\sqrt{2\alpha'}}\left( \alpha_0^\mu+{\widetilde\alpha}_0^\mu
\right)~.
\eeq
In the uncompactified case the two zero modes must be identified because the
string coordinate must be invariant under $\sigma\rightarrow\sigma+\pi$ and
the expression for the momentum in eq.(\ref{gmom}) reduces to the one
obtained
just after eq.(\ref{expc}).

Let us compactify one of the space dimensions along a circle with radius
$R$.
This means that the string coordinate corresponding to this direction that
we
denote for simplicity with $X$ without any index must be periodically
identified as:
\beq
\label{torcomp}
X \sim X +2\pi R~.
\eeq
As in the point particle case, the conjugate momentum corresponding to the
compactified direction must be quantized as
\beq
\label{qmom}
p =\frac{n }{R}~~~~{\rm with}~~~~n \in Z~.
\eeq
This is  simply a consequence of the fact that the generator of the
translations
along the compact direction $e^{ip a }$ must reduce to the identity for
$a =2\pi R$. Moreover in the compactified case the string coordinate $X$
must be
invariant under $\sigma\rightarrow\sigma+\pi$ apart from a factor $2 \pi R
w$
($w$ is an integer) as follows from eq.(\ref{torcomp}). This implies that
\beq
\label{compc}
\pi\sqrt{2\alpha'}
\left( {\alpha}_0 -\widetilde\alpha_0 \right)=2\pi w R~~~~~{\rm with}~~~~~
w \in Z~,
\eeq
where $w$ corresponds to the number of times that the closed string winds
around the compact direction.

Eqs. (\ref{gmom}) and (\ref{qmom}) together with eq. (\ref{compc})
imply that the zero modes  for  the compact direction
must have the following expression
\beq
\label{coscl}
\alpha_0=\sqrt{\frac{\alpha'}{2}}\left(\frac{n}{R}+
\frac{w R}{\alpha'}
\right)~~~~~{\rm and}~~~~~{\widetilde\alpha}_0
=\sqrt{\frac{\alpha'}{2 }}\left(\frac{n}{R}-
\frac{w  R }{\alpha'}\right)~.
\eeq
Inserting eq.(\ref{coscl})  in eq.(\ref{L0c}) and writing also the
contribution
of the uncompactified directions we get
\beq
L_0 =  \frac{\alpha'}{4} {\hat{p}}^2 + \frac{1}{2} \alpha_{0}^{2} +
\sum_{n=1}^{\infty}  \alpha_{-n} \cdot \alpha_n =
\frac{\alpha '}{4} {\hat{p}}^2 +
\frac{\alpha'}{4 }\left(\frac{n }{R}+
\frac{w R }{\alpha'}\right)^2
+
\sum_{n=1}^{\infty}  \alpha_{-n} \cdot \alpha_n
\label{cL0c1}~,
\eeq
and
\beq
\label {cL0c}
\widetilde L_0 = \frac{\alpha'}{4} {\hat{p}}^2 +
\frac {\alpha'}{4 }\left(\frac{n}{R}-\frac{w R }{\alpha'}\right)^2
+
\sum_{n=1}^{\infty}  \widetilde\alpha_{-n} \cdot \widetilde \alpha_n~,
\eeq
The mass operator becomes
\beq
\label{cmassa}
{ M}^2=\frac{2}{\alpha'}
\left[\sum_{n=1}^{\infty}  \left(\alpha_{-n} \cdot \alpha_n+
\widetilde\alpha_{-n} \cdot \widetilde \alpha_n
\right)-2 \right]
+ \left(\frac{n}{R}\right)^2+ \left(\frac{w R}{\alpha'}\right)^2 ~.
\eeq
From the previous expression we see that the spectrum of the closed string
has been enriched by the
appearance of two kinds of particles: the usual K-K
modes which contribute to the energy with $ \frac{n}{R}$
together with some new excitations that are called winding modes
because they can be thought of as generated by the winding of the closed
string around the compact direction which in fact contributes to the energy
of the system as
\beq
\label{winen}
T~ 2\pi R w=\frac{wR}{\alpha'}~,
\eeq
where $T = 1/(2 \pi \alpha')$ is the string tension. All previous
formulas
can be trivially generalized to the case of a toroidal compactified
theory in which more then one coordinate $X^\ell$ is compactified on
circles with radii $R^{(\ell)}.$
Of course in this case we will have K-K and winding modes
corresponding to all the compactified directions.

From eq.(\ref{cmassa}) we see that the spectrum of the theory is
invariant under the exchange of KK modes with winding modes together with
an inversion of the radius of compactification:
\beq
\label{tdua}
w \leftrightarrow n~~~~~~~~~~;~~~~~~~~~~R\leftrightarrow
{\hat R }\equiv\frac{\alpha'}{R}~.
\eeq
This is called a T-duality transformation and $\hat R$ is the
compactification radius of the T-dual theory.  It can also be shown that both
the partition function and the correlators are invariant under T-duality.
This means that
 T-duality  is a symmetry of the bosonic closed string theory.
As a consequence of this invariance, whenever we have to consider
compactified  theories, we can limit ourselves to  the case
$R\geq\sqrt{\alpha'}.$ That is the reason why   $\sqrt{\alpha'}$
is often called the  minimal length of the string theory.

Substituting eq.(\ref{tdua}) into  eq.(\ref{coscl}) we obtain the action
of T-duality on the zero modes
\beq
\label{tduazm}
\alpha_0\rightarrow \alpha_0~~~~~~~~~~;~~~~~~~~~~
\widetilde\alpha_0\rightarrow -\widetilde\alpha_0~.
\eeq
This transformation, however, changes the operators ${\widetilde{L}}_n$
in eq.(\ref{vgenc}) and then does not leave invariant
the physical subspace. In
order to keep the physical subspace invariant we must extend the previous
transformation property to all the non-zero oscillators:
\beq
\label{tduanzm}
\alpha_n\rightarrow \alpha_n~~~~~~~~~~;~~~~~~~~~~
\widetilde\alpha_n\rightarrow -\widetilde\alpha_n~~~~~\forall n\in Z~.
\eeq
This transformation leaves obviously also
invariant the mass of the states given in eq.(\ref{cmassa}).

Eqs.(\ref{tduazm}) and (\ref{tduanzm}) allow us to define the action of
T-duality directly on the string coordinate $X$.
In fact writing
\beq
\label{embleri}
X =\frac{1}{2}\left( X_- + X_+ \right),
\eeq
where
\beq
\label{xmur}
X_- = q + 2 \sqrt{2\alpha'}(\tau-\sigma)\alpha_0
+i\sqrt{2 \alpha'}\sum_{n\neq 0}\frac{\alpha_n}{n}
~e^{-2in(\tau-\sigma)}~~,
\eeq
and
\beq
\label{xmul}
X_+ = q + 2\sqrt{2\alpha'}(\tau+\sigma)\widetilde\alpha_0
+i\sqrt{ 2 \alpha'}\sum_{n\neq 0}\frac{\widetilde\alpha_n}{n}
~e^{-2in(\tau+\sigma)}~~,
\eeq
we see from eqs. (\ref{tduazm}) and (\ref{tduanzm}) that
the T-dual coordinate ${\hat X}$
must satisfy the conditions
\beq
\label{deduts}
\partial_\tau X \rightarrow \partial_{\tau} {\hat{X}} = 
-\partial_\sigma  X~~~~~~~;~~~~~~~
\partial_\sigma X \rightarrow \partial_{\sigma} {\hat{X}} = 
-\partial_\tau  X.
\eeq
They are satisfied if the T-dual coordinate is equal to
\beq
\label{ducor2}
{\hat X} = \frac{1}{2} \left( X_- -X_+ \right)~.
\eeq
Therefore the T-duality transformation acts on the right sector as a parity
operator changing sign of the right moving coordinate $X_+$
and leaving unchanged the left moving one $X_-$.

In an open string theory the string coordinate does not satisfy any
periodicity requirement on $\sigma.$ This implies that in its compactified
version there are only K-K modes, while the winding modes are absent.
This could suggest that T-duality is not a symmetry of the open
string theory. Such a conclusion, however, leads to some problem when
we
remember that  theories with open strings also contain closed strings.
Let us consider a theory with open and closed strings with $d-p-1$
directions
compactified on circles with radii $R^{\ell}$ and take the limit
\beq
\label{dimred}
R^{\ell}\rightarrow 0~~~~~\forall~ {\rm compact~ direction}~,
\eeq
In this limit the open string theory loses effectively $d-p-1$ directions
because all the K-K modes become infinitely massive decoupling from the
spectrum  and because there cannot be any
open string oscillation along the directions with zero radii. Therefore
in this limit the open string will appear to only be living in a
$p+1$-dimensional subspace of the entire $d$-dimensional target space.
Let us analyze what happens in the same limit in the closed string sector.
When the radii of the compact dimensions are vanishing the K-K modes again
decouple, while the winding modes will appear as a continuum of states.
This is a first indication
that for closed strings the compact directions do not disappear from the
theory
as it happened for open strings! More precisely, in the closed string sector
we can perform a T-duality transformation, that is allowed because
it is a symmetry of this sector, and in so doing we can completely
restore all the $d$ space-time dimensions, as a consequence of the fact
that in the limit in eq.(\ref{dimred}) the T-dual radii go to infinity.
But in this way we would end up with a theory in which open strings live in a
$p+1$-dimensional subspace of the entire space-time, while closed strings
live in the entire $d$-dimensional target space. This mismatch can be solved
by requiring that, in the T-dual picture,  open string still can
oscillate in $d$ dimensions, while their endpoints are fixed on a
$p+1$-dimensional hyperplane that we call {\dpb}. Open strings with their
endpoints fixed on these hyperplanes satisfy Dirichlet boundary conditions
in
the $d-p-1$ transverse directions. They are allowed boundary conditions
as we have already seen in eq.(\ref{neudic}) although they destroy the 
Poincar{\`{e}} invariance of the theory. 

In conclusion, in order to avoid a different behaviour between
the closed and the open sector of a string theory,
we must require that the action of T-duality on an open string theory
consists in transforming Neumann boundary conditions into Dirichlet ones.
This
can, in fact, be very naturally obtained if we extend the definition of the
T-dual coordinate given in eq.(\ref{ducor2}) 
to the open string case. In this way we obtain the following T-dual open
string
coordinate:
\beq
\label{ducorr}
{\hat X}^{\ell} = \frac{1}{2}\left[ X_{-}^{\ell} -X_{+}^{\ell} \right]~,
\eeq
where now the left and right movers contain the same set of oscillators
\beq
\label{xmuro}
X_{-}^{\ell} = q^\ell +  c^\ell  
+\sqrt{ 2 \alpha'}(\tau-\sigma)\alpha^\ell_0
+i\sqrt{2\alpha'}\sum_{n\neq 0}\frac{\alpha^\ell_n}{n}
~e^{-in(\tau-\sigma)}~~,
\eeq
and
\beq
\label{xmulo}
X_{+}^{\ell} = q^\ell - c^\ell +
\sqrt{2 \alpha'}(\tau+\sigma)\alpha^\ell_0
+i\sqrt{2\alpha'}\sum_{n\neq 0}\frac{\alpha^\ell_n}{n}
~e^{-in(\tau+\sigma)}~~,
\eeq
From eqs.(\ref{ducorr}), (\ref{xmuro}) and (\ref{xmulo}) one can immediately 
see that T-duality has transformed a string coordinate satisfying
Neumann boundary conditions and given by $1/2 \left[X_{-}^{\ell} + X_{+}^{\ell}
 \right]$ into a  T-dual one satisfying Dirichlet boundary conditions and given
in eq.(\ref{ducorr}). Of course it is also true that, if we had
started with a string coordinate satisfying Dirichlet boundary conditions we
would have obtained a T-dual coordinate satisfying Neumann ones.

The fact that open strings satisfy Dirichlet boundary conditions implies the
existence in the theory of objects, called the {\dpbs}, that are
characterized by the fact that open strings have their endpoints attached
to them. From the three previous equations it also follows that
\beq
\label{onebra}
{\hat X}^{\ell} (\pi)-{\hat X}^{\ell} (0)= - 2\pi \alpha'p^{\ell} = -
\frac{2\pi \alpha' n^{(\ell)}}{R^{(\ell)}}
=- 2\pi n^{\ell} {\hat R^{(\ell)}}
\Rightarrow {\hat X}^{\ell}(\pi)\sim {\hat X}^{\ell} (0)~,
\eeq
This means that in the T-dual theory
the two endpoints of the open string are attached to the same
D-brane.

From the previous construction it also follows that each {\dpb} can be
transformed into a D$p'$-brane through an appropriate sequence of
compactifications and  T-duality transformations.
In fact let us start with a {\dpb} embedded in a $d$-dimensional space-time
and let us compactify one of the space directions $X^\alpha$ that lies in
its
world-volume   on a circle with radius $R^{(\alpha)}.$
The action of a T-duality transformation on this coordinate has the effect
that an open string attached to the brane changes from Neumann to Dirichlet
boundary conditions in that direction. Therefore, the brane 'loses'
one longitudinal coordinate which becomes transverse and is transformed  into 
a D$(p-1)$-brane embedded into a space-time which has the
direction ${\hat X}^\alpha$ compactified on a circle of radius
${\hat R}^{(\alpha)}=\alpha'/R^{(\alpha)}.$ To obtain a
D$(p-1)$-brane living in  the $d$-dimensional  uncompactified space-time  we
need to make the decompactification limit in the T-dual theory, namely
\beq
\label{dc}
{\hat R}^{(\alpha)}\rightarrow\infty,~~{\rm or~equivalently}~~
R^{(\alpha)}\rightarrow 0~.
\eeq
In the same way, if instead of compactifying one of the space-time
directions which are longitudinal to the {\dpb}, we compactify one of the
directions transverse to the brane, and then we act with a T-duality
transformation on this coordinate we will get a  D$(p+1)$-brane
embedded into a space-time with one compact direction.
Then taking the limit in eq.(\ref{dc}) we get again the uncompactified
theory.

To conclude we observe that open strings satisfying Neumann boundary
conditions in all the directions can be thought as being attached to a
space-filling brane that is a D$25$-brane
in the bosonic string or a  D$9$-brane in the superstring.
Therefore, as a consequence of the previous discussion, starting with a
space-filling brane
through a T-duality transformation we can obtain an
arbitrary {\dpb}. More precisely, a Dp-brane can be obtained from a
space-filling brane by first compactifying $d-p-1$ directions, then
performing a T-duality transformation
and finally taking  the decompactification limit.

Up to now we have treated a {\dpb} as a pure geometrical hyperplane to which
open strings are attached and we have completely disregarded the excitations
of the attached open strings. But we will see that, as soon as we let them
come into play, they provide dynamical degrees of freedom to the {\dpb}.

Among all possible excitations of an open string the massless ones have the
peculiarity of not changing the energy of the {\dpb} to which the open
string
is attached. Therefore from the brane point of view they can be interpreted
as collective coordinates of the brane.

In absence of Chan-Paton factors, the massless excitations of an open
string with Neumann boundary conditions in all directions
are described by a $d$-dimensional  abelian gauge potential $A^\mu.$
From the previous discussion  it can be thought as
a gauge  field living on the space-filling brane. Then by
compactifying $d_\perp=d-p-1$ dimensions and making a T-duality
transformation
in each of the compact directions, followed by a decompactification limit in
the T-dual theory, we see that  the vector potential  ${\hat{A}}^\mu$  splits
in a $(p+1)$-dimensional vector ${\hat{A}}^\alpha$ with $\alpha\in\{0,...,p\}$
and $d_\perp$ scalars fields.
The most natural interpretation of the T-dual version of the abelian
gauge field is that the longitudinal coordinates ${\hat{A}}^\alpha$
still describe a gauge field living on the {\dpb} while the
$d_\perp$ scalars coming from the transverse components ${\hat{A}}^{\ell}$,
with $\ell\in\{p+1,...,d-1\}$, appear as the transverse coordinates of the
{\dpb}.

This interpretation becomes more
clear as soon as we introduce a  non-abelian $U(N)$ gauge group
in the open string  theory through the Chan-Paton factors
and in addition we turn on  Wilson lines.
The Chan-Paton procedure for introducing non-abelian gauge degrees of
freedom
on an open string consists in adding  non-dynamical degrees of
freedom at each of its two endpoints. 
A generic string state will therefore be denoted with a ket
$|\alpha,i,\bar j\rangle $
where $\alpha$ describes the usual degrees of freedom of a string,
while the indices $i$ and $\bar j$ refer to the gauge degrees of freedom.
In the case of a $U(N)$ gauge group  $i$ transforms according to 
the fundamental representation $N$,
while $\bar j$ according to the complex conjugate representation $\bar N$ :
\beq
\label{nbarn}
|i'\rangle =U_{i'i}|i\rangle ~~~~~;~~~~~
|\bar j'\rangle =|\bar j\rangle U^+_{jj'}~.
\eeq
If we now introduce a basis of $N\times N$ matrices $\lambda^a_{ij}$, expand 
an open string state as
\beq
\label{wfcp}
|\alpha,a\rangle =\sum_{i,j=1}^N|\alpha,i,\bar j\rangle \lambda^a_{ij}~,
\eeq
and use eq. (\ref{nbarn}) we see that the transformations in
eq.(\ref{nbarn})
act on the matrices $\lambda^a_{ij}$ as follows
\beq
\label{unla}
\lambda^a_{ij}\rightarrow U_{i'i}\lambda^a_{ij} U^+_{jj'}=
\left(U\lambda^a U^+\right)_{i'j'}~.
\eeq
This means that  the matrix $\lambda^a_{ij}$ transforms according to
the adjoint representation ($N\times\bar N$)
of  $U(N)$  which is in fact the appropriate representation for a gauge
field.

Let us  now consider the effect of compactification
in the presence of Chan-Paton factors. For the sake of simplicity
we compactify just one coordinate  that we denote with $X$ without any index
and turn on a pure gauge field of the form
\beq
\label{wlla}
(A  )_{ij}=\frac{1}{2\pi R} {\rm diag}(\theta_1,...,\theta_N)~.
\eeq
This corresponds to a pure gauge configuration generated by the matrix:
\beq
\label{wllu}
U={\rm diag}\left(e^{ iX \frac{\theta_1}{2\pi R}} ,...,
e^{ iX \frac{\theta_N}{2\pi R}}\right)~,
\eeq
because the gauge field configuration in eq.(\ref{wlla}) can be written as
\beq
\label{wllau}
A = - iU^{-1}\partial  U~.
\eeq
But in the case of a compact coordinate the presence of a pure gauge field
affects the parallel  transport along the compact dimension and we get
non-zero Wilson lines:
\beq
\label{wwl}
e^{i\int_0^{2\pi R}A  dx }=
{\rm diag}(e^{i\theta_1},...,
e^{i\theta_N})~.
\eeq
In particular the parallel transport around the compact coordinate transforms
$| i\rangle $ and $|\bar j \rangle $ as follows
\beq
\label{qfc}
|i\rangle \rightarrow e^{i\theta_i}|i\rangle ~~~~;~~~~
|\bar j\rangle \rightarrow e^{-i\theta_j}|\bar j\rangle ~,
\eeq
and therefore the open string state transforms as
\beq
\label{qwf}
|\alpha,a\rangle =\sum_{i,j=1}^Ne^{i(\theta_i-\theta_j)} |\alpha,i,\bar
j\rangle \lambda^a_{ij}~.
\eeq
The presence of Wilson lines changes the possible values of the momentum of
the state $|\alpha, i, \bar{j} \rangle $. In fact in this case a translation
of $2 \pi R$ acts both on the string and the gauge degrees of freedom that are
located at its endpoints. Requiring that this combined action
leaves the state invariant:
\beq
e^{2\pi i R{\hat p}} e^{i(\theta_i-\theta_j)} |\alpha, i, \bar{j} \rangle
= |\alpha, i, \bar{j} \rangle ~,
\label{wtra}
\eeq
we get that the momentum of the state, that we call $p$, is equal to
\beq
\label{wmom}
p =\frac{n}{R}+\frac{(\theta_j -\theta_i)}{2\pi R}~.
\eeq
Let us now see what are the consequences of the presence of  Wilson lines in
the
T-dual theory. Inserting eq.(\ref{wmom}) in eq.(\ref{onebra}) we get
\beq
\label{duebra}
\hat X (\pi)- \hat X (0) = - (2\pi n + \theta_j -\theta_i )\hat R \sim
- (\theta_j -\theta_i)\hat R~.
\eeq
This means that the open string is stretching between two {\dpbs}
whose coordinates are $\theta_i\hat R$ and $\theta_j\hat R.$
Moreover remembering eq.(\ref{wlla}) we immediately see that
\beq
\label{xtrapo}
\theta_i\hat R=2\pi\alpha' (A )_{ii}
~~~~;~~~~
\theta_j\hat R=2\pi\alpha' (A )_{jj}~.
\eeq
and we can conclude that turning on  $U(N)$ Wilson lines in a theory of open
strings along a compactified direction corresponds, in the T-dual theory, to
introduce $N$ {\dpbs} located respectively at
\beq
\label{nbra}
X_1= - 2\pi\alpha' (A )_{11},...,X_N= - 2\pi\alpha' (A )_{NN}
\eeq
In this way the transverse components of a $U(N)$ gauge field carried by an
open string are correctly interpreted as the coordinates of $N$ Dp-branes.

In superstring theory the effect of  T-duality on
the bosonic coordinates is exactly the same as discussed for the bosonic
string, namely  T-duality acts as a parity transformation over the tilded
 sector
\beq
\label{boscor}
X= \frac{1}{2} \left( X_- +X_+ \right) \rightarrow \hat X  = \frac{1}{2} 
\left( X_- -X_+ \right)~.
\eeq
For the fermionic coordinates the transformations under T-duality 
can be fixed by requiring the superconformal invariance of the theory
which imposes
\beq
\label{tduafer}
\psi_+\rightarrow -\psi_+~~~~;~~~~\psi_-\rightarrow \psi_-~,
\eeq
or in terms of the oscillators
\beq
\label{tduafero}
{\widetilde{\psi}}_t\rightarrow -\widetilde\psi_t~~~~;~~~~\psi_t\rightarrow
\psi_t~,
\eeq
This transformation propriety of the fermionic coordinates can also be
understood as due to the requirement that the subspace of the physical
states of the superstring, defined by eqs.(\ref{sphcond1}) and by the
analogous
ones for the right sector, is left invariant by  T-duality.
Therefore, looking at the structure of the operator $\widetilde G_t$ given
in
eq.(\ref{sucurfu})  and taking into account eq.(\ref{tduanzm}), we obtain
again
eq.(\ref{tduafero}).

\section{Classical Solutions Of The Low-Energy String Effective Action}
\label{sec:fifth}

In the previous sections we have seen that the requirement of invariance 
under T-duality transformations in presence of open strings implies the
existence of $p$-dimensional objects called D$p$-branes to which open 
strings can be attached determining their dynamics. Although these objects
are required by T-duality their meaning is still rather obscure in the 
present framework. On the other hand, following a completely different line
of research with the aim to get some non-perturbative information about string
theories  some people were investigating classical solutions of the low-energy
string effective action. The underlying idea was in fact that, as the
construction of 't Hooft-Polyakov monopoles in non abelian gauge theories 
teaches us many things about the non-perturbative structure of non-abelian
gauge theories, so from the study of classical solutions of the low-energy
string effective action one could learn a great deal on non-perturbative
aspects of string theories. It turns out that starting from the low-energy
string effective action one finds solutions of the classical equations of 
motion corresponding to $p$-dimensional objects. In the
following we just want to remind their main properties.

The starting point is the low-energy string effective action containing the
graviton, the dilaton and only one $n$-form potential, that written in the 
Einstein frame is given by:
\beq
S = \frac{1}{2 \kappa^2} \int d^d x \sqrt{-g}  \left[ R  -
\frac{1}{2} \left(\nabla \varphi \right)^2 -
\frac{1}{2 (n+1)!} {\rm e}^{- a \varphi }
\left( F_{n+1} \right)^{2} \right]~~,
\label{action}
\eeq
where $n=p+1$ and $F_{n+1} = d {\cal{C}}$.
For simplicity we have neglected all fermionic fields and the other NS-NS 
and R-R fields. An electric D$p$-brane corresponds to the following
ansatz:
\beq
d {{s}}^2 = \left[ H(r) \right]^{2 A}
\left(\eta_{\alpha\beta}dx^\a
dx^\b \right) + \left[ H(r) \right]^{2B} (\delta_{ij}d x^idx^j)~~,
\label{metri}
\eeq
for the metric ${ g}$, and
\beq
{\rm e}^{- {{\varphi}}(x)} = \left[ H(r) \right]^{\tau}
~~~~,~~~~{\cal{C}}_{01...p}(x) = \pm \sqrt{2 \sigma} [ H(r) ]^{-1}~~,
\label{dil}
\eeq
for the dilaton $ \varphi$ and for the R-R $(p+1)$-form potential
${\cal C}$ respectively. The two signs in ${\cal{C}}$ correspond to
the brane and the anti-brane case and $H (r)$ is assumed to be
only a function of the square of the transverse coordinates $r = x_{\bot}^{2}=
x_i x^i$.  If the parameters are chosen as
\beq
A= - \frac{d - p -3}{2(d-2)}~~~,~~~B= \frac{p+1}{2(d-2)}~~~,~~~
\tau= \frac{{{a}}}{2}~~~,~~~\sigma = \frac{1}{2}~~,
\label{expo}
\eeq
with ${{a}}$ obeying the equation
\beq
\frac{2 (p+1)( d-p-3)}{d-2}  + {{a}}^2 = 4~~,
\label{equ}
\eeq
then the function $H(r)$ satisfies the flat space Laplace equation.
An extremal $p$-brane solution is constructed by introducing in the
right hand side of the eqs. of motion following from the action in 
eq.(\ref{action}) a $\delta$-function source
term in the transverse directions.
If we restrict ourselves to the simplest case of just one
$p$-brane, we obtain the following expression for $H (r)$
\beq
H(r) = 1 + 2 \kappa T_p G(r) ~~,
\label{sol}
\eeq
where
\beq
G(r) = \left\{ \begin{array}{cc}
 \left[ (d-p -3) r^{(d-p-3)}
\Omega_{d-p-2}\right]^{-1} &  ~~~p < d-3 ~~,\\
  - \frac{1}{2 \pi} \log r  &   ~~~p= d-3 ~~,
\end{array} \right.
\label{solu}
\eeq
with 
\beq
\Omega_{q}={ 2 \pi^{(q+1)/2}}/{ \Gamma \Big((q+1)/2\Big)}
\label{in8}
\eeq
being the area of a unit $q$-dimensional sphere $S_q$. 
For future use it is convenient to introduce the quantity:
\begin{equation}
Q_p = \mu_p\,\frac{\sqrt{2}\,\kappa\,}{(d-p-3)\,\Omega_{d-p-2}}~~;~~
\mu_p \equiv \sqrt{2} T_p~~.
\label{Qp}
\end{equation}
and (if $p < d-3$) to rewrite $H(r)$ in eq.(\ref{sol}) as follows:
\beq
H(x) = 1 +  \frac{Q_p}{r^{d-3-p}}~~,
\label{conv43}
\eeq
The classical solution has a mass per unit $p$-volume, $M_p$ and an electric
charge with respect to the R-R field, $\mu_p$, given respectively by
\beq
M_p = \frac{T_p}{\kappa}~,~ \hspace{1cm} \mu_p = \pm \sqrt{2} T_p~~.
\label{char}
\eeq
The fundamental observation made by Polchinski~\cite{POL95} has been to 
identify the 
D$p$-branes required by T-duality with the $p$-branes obtained as classical
solutions of the low-energy string effective action. Therefore, on the one hand
the $p$-branes are new non-perturbative states of string theory and on the
other hand have the important property that open string have their endpoints
attached on them. The latter property will allow one to compute their 
interactions
and more in general to study their properties by computing open string one-loop
diagrams. On the other hand we should not be worried that the Dirichlet boundary
conditions break  Poincar{\`{e}} invariance because this 
happens in presence of any kind of solitonic state. In the next chapter we will
introduce the boundary state and we will show that it provides a stringy 
description of these new states. 

\section{Bosonic Boundary State}
\label{sec:sixth}

As discussed in the previous section 
D$p$-branes  are extended $p$ dimensional objects  characterized by
the fact that open strings can have their endpoints attached to them.
In general they are dynamical and non rigid objects, that can fluctuate in
shape and on which external fields can live. In these lectures we limit
ourselves to treat them as static and rigid objects.

The open string with the endpoint at $\sigma=0$ attached to a D$p$-brane
satisfies the usual Neumann boundary conditions along the directions
longitudinal to the world volume of the brane
\beq
\partial_{\sigma} X^{\alpha}|_{\sigma=0} =0 \hspace{2cm} \alpha=0, 1, ....,
p
\label{neu1}~,
\eeq
and  Dirichlet boundary conditions along the directions transverse to the
brane
\beq
X^{i} |_{\sigma=0} = y^i   \hspace{2cm} i= p+1, ....d-1
\label{dir1}~,
\eeq
where $y^i$ are the coordinates of the brane and $d$ is the dimension of the
Minkowski space-time, that in the case of the bosonic string is equal to
$d=26$.

As the interaction between two superconducting plates is obtained by
computing the vacuum fluctuation of the electromagnetic field that gives rise to
the Casimir effect, so the interaction between  two D$p$-branes is given by 
the vacuum fluctuation of an open string that is
stretching between them. 
This means that their interaction
is simply given by the one-loop open string "free-energy"
which is usually represented by the annulus  or equivalently
by the cylinder diagram.
From either of those two diagrams it is easy to see that by exchanging  
the variables $\sigma$ and $\tau$
the one-loop open string amplitude can also be viewed as a tree diagram of a
closed string created from the vacuum, propagating for a while and then
annihilating again into the vacuum. These two equivalent descriptions of
the same diagram are called respectively the `open-channel' and
the `closed-channel'.
We want to stress that the physical content of the two descriptions is
a priori completely different. In the first case we describe the interaction
between two {\dpbs} as a one-loop amplitude of open strings, which is the
amplitude of a quantum theory of open strings, while
in the second case we describe the same interaction as
a tree-level amplitude of closed strings, which is instead a classical
amplitude in a theory of closed strings.
The fact that these two descriptions are equivalent is a consequence
of the conformal symmetry of string theory that allows one to
connect the two apriori different descriptions.

To show that, let us consider a one-loop diagram with  an open string
circulating in it and stretching between two parallel D$p$-branes
with coordinates respectively  given by $(y^{p+1},...,y^{d-1})$
and $(w^{p+1},...,w^{d-1})$. The open string satisfies the 
boundary conditions in eq.(\ref{neu1}) both at $\sigma=0 $ and $\sigma = \pi$
along the
world-volume directions of the brane, while along the transverse directions
satisfies the following equations:
\beq
\label {bc2}
X^i|_{\sigma=0}=y^i~~~~~~~~~~ X^i|_{\sigma=\pi}=w^i~~~~~
i = p+1,..., d-1~,
\eeq
where we take $\sigma$ and $\tau$ in the two intervals $\sigma\in[0,\pi]$
and
$\tau\in [0, T].$

We now want to find a  conformal transformation acting on the previous open
string boundary conditions in order to transform them into the boundary
conditions for a closed string propagating between the two D$p$-branes.
In terms of the complex coordinate  $\zeta\equiv\sigma+i\tau,$
a conformal transformation simply transforms $\zeta\rightarrow f(\zeta),$
where $f(\zeta)$ is an arbitrary holomorphic function of $\zeta.$
Let us consider the following conformal transformation
\beq
\label {ctz}
\zeta = \sigma + i \tau \rightarrow -i\zeta=\tau - i\sigma~.
\eeq
After the inversion $\sigma\rightarrow -\sigma$ the previous conformal
transformation
simply amounts to exchange $\sigma$ with $\tau$ and viceversa
\beq
\label{ctst}
(\sigma, \tau)\rightarrow (\tau, \sigma)~.
\eeq
Finally in order to have the closed string variables $\sigma$ and $\tau$ to
vary in the  intervals $\sigma\in[0, \pi]$ and  $\tau \in [0, \hat{T}]$
corresponding to a closed string propagating between the two D branes one
must
perform the following conformal rescaling
\beq
\label{res}
\sigma \rightarrow \frac{\pi}{T}\sigma ~~~~~~~~~~
\tau\rightarrow \frac{\pi}{T}\tau~,
\eeq
with  $\hat{T}=\pi^2/T.$ We have therefore constructed a conformal
transformation that brings us
from the open string to the closed string channel.
In the closed string channel we need to construct the two boundary
states $|B_X \rangle $ that
describe the two D$p$-branes respectively at $\tau=0$ and $\tau ={\hat{T}}$.
The equations that characterize these states are obtained
by applying the conformal transformation previously constructed to the
boundary conditions for the open string given in eqs.(\ref{neu1}) and
(\ref{bc2}). At $\tau =0$ we get the following conditions:
\begin{equation}
\label {bc1c}
\partial_{\tau}X^\alpha|_{\tau=0}|B_X \rangle =0 ~~~~~~~~~~\alpha =0,...,p~,
\end{equation}
\begin{equation}
\label {bc2c}
X^i|_{\tau=0}|B_X \rangle =y^i ~~~~~~~~~~i = p+1,..., d-1~.
\end{equation}
Analogous conditions can be obtained for the D$p$-brane at
$\tau = {\hat{T}}$.

The previous equations can be easily written in terms of the closed string
oscillators by making use of the expansion in eq.(\ref{expc}), obtaining
\beqa
\label{over1}
(\alpha_n^\alpha+\widetilde\alpha_{-n}^\alpha)|B_X \rangle  =0~~;~~
(\alpha_n^i-\widetilde\alpha_{-n}^i)|B_X \rangle  =0~~\forall n\neq 0
\nonumber\\
{\hat{p}}^\alpha|B_X \rangle  = 0 ~~~~~~~~~~({\hat{q}}^i-y^i)|B_X \rangle =0~.
\eeqa
Introducing the matrix
\beq
\label{matS}
S^{\mu\nu}=(\eta^{\alpha\beta},-\delta^{ij})~,
\eeq
the equations for the non-zero modes can be rewritten as
\beq
\label {over1s}
(\alpha_n^\mu+S^\mu~_\nu\widetilde\alpha_{-n}^\nu)|B_X \rangle
=0~~~~~\forall \,\, n \neq 0
~.
\eeq
The state satisfying the previous equations can easily be determined to be
\beq
\label{b1}
|B_X \rangle  =N_p \delta^{d-p-1}({\hat q}^i-y^i) \left(\prod_{n=1}^\infty
e^{-\frac{1}{n}
\alpha_{-n}
S\cdot\widetilde\alpha_{-n}}\right)|0\rangle _{\alpha}|0\rangle
_{\widetilde\alpha}
|p=0\rangle ~,
\eeq
where $N_p$ is a normalization constant to be fixed.

The previous boundary state describes only the degrees of freedom
corresponding
to the string coordinate $X$. In order to have a BRST invariant boundary
state
we have to supplement it with the boundary state for the ghost degrees of
freedom obtaining the full boundary state
\beq
\label{bmg}
|B \rangle =|B_X \rangle  |B_{gh} \rangle   ~.
\eeq
We will later on write the ghost part of the boundary state.

The overlap conditions for the conjugate boundary state can be easily
obtained by taking the adjoint of eqs. (\ref{over1}) and (\ref{over1s}) and 
are given by
\beq
\label{conj1}
\langle
B_X|\left(\alpha_{-n}^\mu+S^\mu~_\nu\widetilde\alpha^\nu_n\right)~~;~~
\langle B_X|\hat p^\alpha=0~~;~~\langle B_X|(\hat q^j-y^j)=0~~,
\eeq
that imply
\beq
\label{conj2}
\langle B_X|= \langle p=0|~_\alpha\langle 0|~_{\widetilde\alpha}\langle 0|~
N_p \delta^{d-p-1}({\hat q}^i-y^i) \left(\prod_{n=1}^\infty e^{-\frac{1}{n}
\alpha_{n}
S\cdot\widetilde\alpha_{n}}\right)~~,
\eeq

In the following we will compute the interaction between two parallel 
D$p$-branes both in the open and in the closed string channel. By comparing
the two results we can determine the normalization factor $N_p$ appearing
in the boundary state in eq.(\ref{b1}). Let us start by
computing the interaction in the closed string channel.
For the sake of simplicity we
perform this calculation considering only the part of the boundary state
containing the string coordinate $X$ and then adding by hand the
contribution of the ghosts. 
With this simplification the free energy reads as
\begin{equation}
F =  \langle B_X |  D |B_X \rangle
\label{abb}~,
\end{equation}
where $ D$ is the bosonic closed string propagator
\beq
\label{propg}
D= \frac{\alpha'}{4\pi}\int_{|z|\leq 1}\frac{d^2z}{|z|^2}z^{L_0 -1 }\bar
z^{\widetilde L_0 -1 }~,
\eeq
$L_0$ and $\widetilde L_0$ are the usual Virasoro operators of the closed
bosonic string given in eq. (\ref {L0c}). 

Inserting eqs.(\ref{b1}), (\ref{conj2}) and (\ref{propg}) into eq.
(\ref{abb})  we get
\beqa
\langle B_X |D| B_X \rangle=
(N_p)^2\frac{\alpha'}{4\pi}\int_{|z|\leq 1}\frac{d^2z}{|z|^4}
\langle 0|_\alpha\langle 0|_{\widetilde\alpha}\langle p=0|
\prod_{n=1}^\infty \left(e^{- \frac{1}{n}
\alpha_n\cdot S\cdot{\widetilde\alpha}_n} \right)
\nonumber\\
\delta^{d_\perp}({\hat q}_i)~z^{L_0}\bar z^{\widetilde L_0}~
\delta^{d_\perp}({\hat q}_i-y_i)~
\prod_{n=1}^\infty \left( e^{-\frac{1}{n} \alpha_{-n} \cdot
S \cdot {\widetilde{\alpha}}_{-n}} \right)
|0\rangle_{\alpha}|0\rangle_{\widetilde\alpha}|p=0\rangle
~~,
\label{clc}
\eeqa
where  $ d_\perp \equiv d-p-1$ and $|y|$ is the distance between  the two
{\dpbs}.
The matrix element in the previous expression can be factorized in two parts
containing respectively the contribution of the zero and non-zero modes.
The contribution of the zero modes is given by:
\[
\langle p=0|\delta^{d_\perp}(\hat q_i)|z|^{\frac{\alpha'}{2}{\hat p}^2}
\delta^{d_\perp}(\hat q_i-y_i)|p=0\rangle=
\]
\[
= \int\frac{d^{d_\perp}Q}{(2\pi)^{d_\perp}}\int\frac{d^{d_\perp}Q'}
{(2\pi)^{d_\perp}}
\langle p=0|e^{iQ\cdot \hat q} |z|^{\frac{\alpha'}{2}\hat
p^2}e^{iQ'\cdot(\hat q-y)}|p=0 \rangle=
\]
\[
=
\int\frac{d^{d_\perp}Q}{(2\pi)^{d_\perp}}\int\frac{d^{d_\perp}Q'}{(2\pi)^{
d_\perp}}
|z|^{\frac{\alpha'}{2}{Q'}^2}e^{-iQ'\cdot y}\langle p_{\perp}=-Q|{p}_{\perp}
= Q' \rangle
\langle p_{\parallel}=0|{p}_{\parallel}=0\rangle=
\]
\beq
\label{zp}
=V_{p+1}\int\frac{d^{d_\perp}Q}{(2\pi)^{d_\perp}}|z|^{ \frac{\alpha'}{2}{Q}^
2}
e^{ iQ \cdot y}~,
\eeq
where
we have used the following normalization for each component of the momentum
\beq
\label{knorm}
\langle k|k'\rangle=2\pi \delta(k-k')~,
\eeq
with
\beq
\label{vol}
(2\pi)^d\delta^d(0)\equiv V_d~.
\eeq
Performing the gaussian integral, eq. (\ref{zp}) becomes
\beq
\label{zpp}
V_{p+1} e^{-y^{2}/(2 \pi \alpha' t)} \left( 2 \pi^2 t \alpha'
\right)^{-d_{\perp}/2}~~~,~~~ |z| = { e}^{- \pi t }~.
\eeq
The contribution of the non-zero modes is instead given by
\[
{}_\alpha\langle 0| {}_{\widetilde\alpha}\langle 0|~\prod_{m=1}^\infty 
\left( e^{- \frac{1}{m}\alpha_m\cdot S\cdot\widetilde\alpha_m} \right) z^N 
\bar z^{\widetilde N}\prod_{n=1}^\infty
\left( e^{-\frac{1}{n} \alpha_{-n} \cdot
S \cdot {\widetilde{\alpha}}_{-n}} \right) 
|0\rangle_{\alpha}|0\rangle_{\widetilde\alpha}=
\]
\beq
\label{oscp}
={}_\alpha\langle 0| {}_{\widetilde\alpha} \langle 0| ~ \prod_{m=1}^\infty
e^{-\frac{1}{m}
\alpha_m\cdot S\cdot{\widetilde\alpha}_m}\prod_{n=1}^\infty
e^{-\frac{1}{n}
\alpha_{-n}\cdot
S\cdot{\widetilde{\alpha}}_{-n}|z|^{2n}}|0\rangle_{\alpha}|0
\rangle_{\widetilde\alpha}~,
\eeq
where we have defined
\beq
\label{N}
N\equiv\sum_{n=1}^{\infty}  \alpha_{-n} \cdot \alpha_n~~~~~;~~~~~
\widetilde N\equiv\sum_{n=1}^{\infty}  {\widetilde\alpha}_{-n} \cdot
{\widetilde\alpha}_n~,
\eeq
and we have used the following relations:
\beq
\label{zN}
z^N e^{\alpha_{-n}}z^{-N}=e^{\alpha_{-n} z^n}~~~~~{\rm and}~~~~~
{\bar z}^N e^{\alpha_{-n}}{\bar z}^{-N}=
e^{\alpha_{-n} {\bar z}^n}~~~~~\forall n
\neq 0~.
\eeq
By explicitly evaluating the contractions among the oscillators in
eq.(\ref{oscp}) we get
\beq
\label{ser}
\prod_{n=1}^\infty\left(\frac{1}{1-|z|^{2n}}\right)^{d-2}~.
\eeq
To be more precise the previous calculation leads to a power $d$ instead of
$d-2$ as we have written in eq.(\ref{ser}). The extra power $(-2)$ comes
from
the ghost contribution that, for the sake of simplicity, we are not
presenting
here. 

Inserting eqs.(\ref {zp}) and (\ref {ser}) in eq.(\ref {clc})
after having changed variables according to 
\beq
\label{chva}
|z|=e^{-\pi t}~~~~~d^2z=-\pi e^{-2\pi t}dtd\varphi~,
\eeq
we get
\[
\langle B_X | D|B_X \rangle =
\]
\[
=  (N_p)^2V_{p+1}\frac{\alpha'\pi }{2}\int_{0}^\infty dt~
\left( 2 \pi^2  \alpha' t \right)^{-\frac{d_{\perp}}{2}}
e^{-\frac{y^2}{2\pi\alpha' t}}
e^{2\pi t}
\prod_{n=1}^{\infty} \left(\frac{1}{1-e^{- 2\pi t n}}\right)^{d-2}
=
\]
\beq
\label{intg}
= (N_p)^2 V_{p+1}\frac{\alpha'\pi}{2}
\left( 2 \pi^2  \alpha'\right)^{-\frac{d_\perp}{2}}
\int_{0}^\infty
dt~ t^{-\frac{d_\perp}{2}} ~ e^{-\frac{y^2}{2\pi\alpha' t}}
[f_1(e^{-\pi t})]^{-24}~
\eeq
where we have introduced the function 
\beq
\label{f1}
f_1 (q) \equiv q^{{1\over 12}} \prod_{n=1}^\infty (1 - q^{2n})~~;
\eeq
The factor $\alpha' \pi/2= \alpha'/(4 \pi) \pi (2 \pi)$ in eq.(\ref{intg}) 
comes from the product of the factor $\alpha' /(4 \pi)$ present in the
propagator in eq.(\ref{propg}), the factor $\pi$ obtained in eq.(\ref{chva})
and the factor $2 \pi$ obtained by performing the trivial integration over 
the angular variable $\varphi$.

Let us now proceed to the calculation of the interaction between two 
D$p$-branes in the open string channel. The one-loop planar free-energy for
an open string with $d-p-1$ Dirichlet boundary conditions is equal to
\footnote{Note that here
we use the regularized expression
$\log (x)= - \lim_{\varepsilon\rightarrow 0^+}
\int_\varepsilon^\infty \frac{d \tau}{\tau}e^{- \tau x}$
}
\beq
F = -\frac{1}{2} Tr \log \left[ L_0  -1 \right] =
\int_{0}^{\infty} \frac{d\tau}{2\tau} Tr
\left[ e^{- 2\pi ( L_0  -1 )\tau} \right]~, 
\label{freet}
\eeq
where
\beq
L_0  = \alpha ' k^2 + \frac{y^2}{(2 \pi)^2 \alpha '} + 
\sum_{n=1}^{\infty}  \alpha_{-n} \cdot \alpha_n~,
\label{L0o}
\eeq
$y$ represents the distance between the two parallel D$p$-branes and
$k$  the momentum lying along the world volume of the two branes.
Here the $L_0$ operator  differs from the one in eq. (\ref{L0on})
because in this case
the open string satisfies  Dirichlet boundary conditions 
in the $d-p-1$ transverse directions 
and  Neumann boundary conditions in the longitudinal ones. 
 
The trace in eq.(\ref{freet}) must be understood as an integration 
over the longitudinal loop momentum and a trace over the oscillators, namely
\[
F=2
\frac{V_{p+1}}{2} \int\frac {d^{p+1} k}{(2\pi)^{p+1}} \times
\]
\[
\times \int_0^\infty \frac {d\tau}{\tau}
~ e^{2\pi \tau}~ e^{-2\pi \tau\alpha' k^2} ~e^{- \frac{y^2 \tau}{2\pi\alpha'}}~
{\rm e}^{2 \pi \tau}~ 
Tr \left(\prod_{n=1}^\infty e^{-2\pi \tau \alpha_{-n} \cdot \alpha_{n}}\right)=
\]
\beq
\label {free1}
=
V_{p+1}\int_0^\infty \frac {d\tau}{\tau} ~
(8\pi^2\alpha'\tau)^{-\frac{p+1}{2}}~
e^{-\frac{y^2 \tau}{2\pi\alpha'}}~
Tr \left(\prod_{n=1}^\infty e^{-
2\pi \tau \alpha_{-n} \cdot \alpha_{n}}\right)~,
\eeq
where we have performed the Gaussian integral over the longitudinal momentum
circulating in the loop and we have
inserted a factor $2$ coming from the freedom of exchanging the
two endpoints of the string.
Evaluating explicitly the trace over the oscillators
we get
\beq
\label{evtr}
Tr \left(\prod_{n=1}^\infty e^{-
2\pi \tau \alpha_{-n} \cdot \alpha_{n}}\right)
=\prod_{n=1}^\infty \left(\frac{1}{1-e^{-2\pi \tau n}}\right)^{d-2}~.
\eeq
Notice that also in this case we have introduced by hand
the information that the ghosts contribution amounts to change the exponent
from $d$ to $d-2$.
Finally, inserting eq. (\ref{evtr}) into eq. (\ref{free1})
and writing it
in terms of the function $f_1$ defined in eq.(\ref{f1}), the one-loop
free-energy becomes
\[
F =
V_{p+1} \, (8\pi^2\alpha')^{-\frac{p+1}{2}} \int_0^\infty 
\frac {d\tau}{\tau} \tau^{-\frac{p+1}{2}}
~e^{-\frac{y^2 \tau}{2\pi\alpha'}}~\left(f_1(e^{-\pi \tau})
\right)^{-24}=
\]
\beq
\label{free2}
=V_{p+1}\, (8\pi^2\alpha')^{-\frac{p+1}{2}}
\int_0^\infty \frac {d\tau}{\tau}~\tau^{12 -\frac{p+1}{2}}
e^{-\frac{y^2 \tau}{2\pi\alpha'}}~\left(f_1(e^{-\frac{\pi}{\tau}}) 
\right)^{-24}~,
\eeq
where we have taken $d=26$ and we have
used the modular transformation property
of the function $f_1$ 
\beq
\label{mtf1}
f_1(e^{-\frac{\pi}{t}})=\sqrt{t}f_1(e^{-\pi t})
\eeq
In order to compare eq.(\ref{free2}) with eq.(\ref{intg}) we must perform
in the second one the change of variable  $t = \frac{1}{\tau}$. In
this way eq.(\ref{intg}) becomes
\[
\langle B_X | D|B_X \rangle=
\]
\beq
\label {csc}
= (N_p)^2 V_{p+1}
\frac{\alpha'\pi}{2} \left( 2 \pi^2 \alpha' \right)^{- \frac{d_\perp}{2}}
\int_{0}^\infty \frac{d\tau}{\tau}~\tau^{12-\frac{p+1}{2}}
e^{-\frac{y^2 \tau}{2\pi\alpha'}} [ f_1(e^{-\frac {\pi} {\tau}})]^{-24}~.
\eeq
By comparing eqs.(\ref{free2}) and (\ref{csc}) we can determine the
normalization factor of the boundary state:
\beq
\label {Tp}
N_p=\frac{T_p}{2}~~~,~~~
T_p=\frac {{\sqrt \pi}}{2^{\frac{d-10}{4}}}
(2\pi\sqrt\alpha')^{\frac{d}{2}-2-p} ~.
\eeq
In conclusion, by performing the calculation of $F$ in the closed string channel
with the normalization factor given in eq.(\ref{Tp}) we get:
\beq
F = V_{p+1} \left( 8 \pi^2 \alpha' \right)^{- \frac{p+1}{2}}\int_{0}^{\infty}
\frac{dt}{t}\,\,t^{\frac{p+1}{2} -12} 
e^{-\frac{y^2 }{2\pi\alpha' t}}~\left(f_1(e^{- \pi t}) 
\right)^{-24}~,
\label{clocha}
\eeq
while performing the calculation in the open string channel we get:
\beq
F= V_{p+1}\,(8\pi^2\alpha')^{- \frac{p+1}{2}}
\int_0^\infty \frac {d\tau}{\tau}~\tau^{- \frac{p+1}{2}}
e^{-\frac{y^2 \tau}{2\pi\alpha'}}~\left(f_1(e^{- \pi \tau}) 
\right)^{-24}~,
\label{opecha}
\eeq
The two expressions are manifestly identical as one can see if one changes 
variable
$\tau = 1/t$ in eq.(\ref{opecha}) and uses the modular transformation in 
eq.(\ref{mtf1}).

Notice that the function $[ f_1]^{-24}$ has the following expansion for large
value of the argument ($x \rightarrow \infty$):
\beq
\left[f_1 ( e^{- \pi x})  \right]^{-24} = 
\sum_{n=0}^{\infty} c_n e^{- 2 \pi x (n-1)} 
= e^{2 \pi x} + 24 + 0(e^{- 2 \pi x})~.
\label{expf1}
\eeq
In the open string channel the $n$th term of the previous expansion corresponds
to the contribution in the loop of open string states with mass 
$\alpha' M^2 = n-1$,
while in the closed string channel corresponds to the exchange between the
two branes of closed string states with mass $\frac{\alpha'}{2} M^2 = 2(n-1)$.
In particular from eq.(\ref{clocha}) we can see that the dominant contribution
to $F$ at large distance ($ y \rightarrow \infty$) comes from light closed 
string states:
\beq
F = V_{p+1} \left( 8 \pi^2 \alpha' \right)^{- \frac{p+1}{2}}\int_{0}^{\infty}
\frac{dt}{t}\,t^{\frac{p+1}{2} -12} 
e^{-\frac{y^2 }{2\pi\alpha' t}}~\left( e^{2 \pi t} + 24 + \dots \right)~,
\label{clochaex}
\eeq
where the first term corresponds to the exchange between the two branes
of the closed string tachyon, the second term to the exchange of 
the massless closed string states and the additional terms to the
exchange of closed string states with higher mass. The first term is obviously
divergent, but it is due to the presence of the tachyon that will disappear 
in superstring. The second term, that is called massless tadpole, 
can be cancelled if we add the contribution of the 
non-orientable Moebius diagram and we choose a particular gauge group 
($SO(2^{13})$ for the bosonic string or $SO(32)$ for the type I superstring).
The requirement of tadpole cancellation is
a convenient way of fixing the gauge group besides the one of
anomaly cancellation. Both ways fix the gauge group in the type I theory 
to be $SO(32)$.

In the field theory limit ($\alpha ' \rightarrow 0$) it is more convenient to 
use the expression for $F$ written in the open string channel because in this 
case the dominant contribution comes from the lowest open string states 
circulating in the loop. This limit can be conveniently done by introducing in 
eq.(\ref{opecha}) the dimensional Schwinger proper time $s$ related to the
modular parameter $\tau$ by the relation $ s = \alpha ' \tau$. Then  using 
the expansion in eq.(\ref{expf1}) we can rewrite eq.(\ref{opecha}) as follows:
\beq
F= V_{p+1}(8\pi^2 )^{- \frac{p+1}{2}}
\int_0^\infty \frac {d s}{s}~s^{- \frac{p+1}{2}}
e^{-\frac{y^2 s}{2\pi (\alpha')^2}}~\left( e^{2 \pi s/\alpha' } + 24 + 
0( e^{-2 \pi s/(\alpha ')^2} \right)~.
\label{opechaex}
\eeq 
The first term corresponds to the open string tachyon that will not be present
in superstring and the second term corresponds to the open-string massless 
states. Finally the additional terms correspond to states with higher mass in 
open string theory that are negligible for $\alpha ' \rightarrow 0$. Notice
that, if we neglect the tachyon contribution that is absent in superstring,
the massless states give a non vanishing contribution to $F$ only if the
distance between the two branes $y \rightarrow 0$.. 

As we have seen in eq.(\ref{bmg}) the BRST invariant boundary state is the
product of the boundary state $|B_X \rangle$ for the bosonic coordinate,
that we have constructed in this section, and of $|B_{gh} \rangle$
that we are now going to construct. BRST invariance requires that the total
boundary state satisfies the equation
\beq
\label{BRST}
(Q+\widetilde Q)|B \rangle =0~,
\eeq  
where the BRST charge, given in eq.(\ref{brchar}), is equal to
\beq
Q = \sum_{n} c_n L_{-n}^{X} + \sum_{n=-1}^{\infty} c_{-n} L_{n}^{gh} +
\sum_{n=2}^{\infty} L^{gh}_{-n} c_n 
\label{brstq}
\eeq
${\tilde{Q}}$ is given by an analogous expression in terms of the tilded 
variables. 
The overlap conditions in eq.(\ref {over1}) imply that the boundary state for
the bosonic coordinate satisfies the following eqs.:
\beq
\label{ovgv}
(L_{m}^{X} - {\widetilde L}_{-m}^{X} )|B_X\rangle =0~.
\eeq
Inserting the expression for $Q$ and the corresponding expression for 
${\tilde{Q}}$ in eq.(\ref{BRST}) and using eq.(\ref{ovgv}) we can 
see that eq.(\ref{BRST}) implies the following overlap conditions for the
ghost boundary state
\beq
\label{overgh}
(c_n+{\widetilde c_{-n}})|B_{gh}\rangle =0~~~~~~;~~~~~~
(b_n-{\widetilde b}_{-n})|B_{gh}\rangle =0~.
\eeq
The second overlap condition in the previous equation follows from the
first one and from the analogous of eq.(\ref{ovgv}) for the ghost boundary 
state:
\beq
\label{ovgvg}
(L_{m}^{gh} - {\widetilde L}_{-m}^{gh} )|B_gh\rangle =0~.
\eeq
Eqs.(\ref{overgh}) are satisfied by the state
\beq
\label{bgh}
|B_{gh}\rangle _{gh}= e^{\sum_{n=1}^\infty (c_{-n}{\widetilde b}_{-n}-
b_{-n}{\widetilde c}_{-n})}
\left(\frac {c_0+{\widetilde c}_0}{2}\right)|q=1\rangle |{\widetilde{q=1}}
\rangle
\eeq
where $|q=1\rangle $ is the state that is annihilated by the following
oscillators
\beq
\label{q1}
c_n|q=1\rangle =0~~~~~\forall n\geq 1;~~~~~;~~~~~~
b_m|q=1\rangle =0~~~~~\forall m\geq 0~.
\eeq

\section{Fermionic Boundary State}
\label{sec:seventh}

In this section we want to generalize the previous construction to the
superstring case where, together with the boundary state $|B_X \rangle $
corresponding
to the bosonic coordinate $X$ that we have already constructed
we also need to determine the boundary state $|B_{\psi} \rangle $ corresponding
to the
fermionic
coordinate $\psi$. The procedure that we follow for determining $|B_{\psi}
\rangle $
is precisely the same used in the previous section. We perform on the 
boundary conditions for an open string stretching between two D$p$-branes
the conformal transformation
that brings from the open string to the closed string channel. In this way
we obtain the equations that the fermionic boundary state must satisfy. We
then
solve them finding the explicit expression for $|B_{\psi} \rangle $.

Let us consider the boundary conditions of an open superstring
stretching between two D$p$-branes and circulating in the planar loop
describing the interaction between  two parallel branes.
If the bosonic degrees of freedom satisfy  Neumann boundary conditions in
all
the directions we have the following boundary conditions for the fermionic
coordinate:
\beq
\label{fbcu}
\left\{
\begin{array}{l}
\psi_-(0,\tau)=\eta_1{\psi_+}(0,\tau)\\
\psi_-(\pi,\tau)=\eta_2{\psi_+}(\pi,\tau)
\end{array}
\right.
\eeq
where $\eta_1$ and $\eta_2$ can take the values $\pm 1$. If $\eta_1 =\eta_2$
we get the Ramond (R) sector, while if $\eta_1 =- \eta_2$ we get instead the
Neveu-Schwarz (NS) sector.

In order to understand how they change when the bosonic coordinate
satisfies Dirichlet boundary conditions in some of the directions, we
compactify them and apply T-duality. A
T-duality transformation along a direction $i$
of an open string theory transforms Neumann into Dirichlet boundary 
conditions for
the bosonic coordinate and, as discussed in 
sect.~\ref{sec:fourth}, changes the sign of the
fermionic coordinate in the right sector leaving that of the left sector
unchanged, i.e.
\beq
\label{tduaf}
\psi^i_{-}\rightarrow\psi^i_{-}~~~~~~~{\rm}~~~~~~~
\psi^i_+\rightarrow -\psi^i_+
\eeq
Therefore the boundary conditions in eq.(\ref{fbcu}) are generalized to the
case of
an open superstring satisfying  Neumann boundary conditions in the
directions
longitudinal to the world-volume of the D$p$-brane and  Dirichelet boundary
conditions in the transverse directions, as follows
\beq
\label{fbcuS}
\left\{
\begin{array}{l}
\psi^\mu_-(0,\tau)=\eta_1 S^{\mu}_{\,\,\,\nu} {{\psi}_+}^{\nu} (0,\tau)\\
\psi^\mu_-(\pi,\tau)=\eta_2 S^{\mu}_{\,\,\, \nu}{{\psi}_+}^{\nu}(\pi, \tau)
\end{array}
\right.
\eeq
where the matrix $S$ has been defined in eq.(\ref{matS}).

But, together with the assignment of the usual boundary conditions
that connect the left and right modes at the endpoints of
the open superstring, we must also give the  periodicity or anti
periodicity conditions for the fermionic degrees of freedom in going around
the loop. These are chosen to be
\beq
\label{fbcp}
\left\{
\begin{array}{l}
\psi_-(\sigma,0)=\eta_3 \psi_-(\sigma, T)\\
{\psi}_+(\sigma,0)=\eta_4{\psi}_+(\sigma, T)
\end{array}
\right.
\eeq
where $\eta_3$ and $\eta_4$ can take the values $\pm 1$. From the boundary
conditions in eqs.(\ref{fbcuS}) and (\ref{fbcp}) we get
\beq
\label {eta34}
\psi^\mu_-(0,0)=\eta_1 S^{\mu}\,_{\nu}\psi^\nu_+(0,0)=
\eta_1 \eta_4 S^{\mu}_{\,\,\, \nu} {{\psi}_+}^\nu (0, T)
\eeq
and
\beq
\label{eta342}
\psi^\mu_-(0,0)=\eta_3  \psi^\mu _-(0, T)= \eta_3 \eta_1
S^{\mu}_{\,\,\, \nu} {{\psi}_+}^\nu(0, T)
\eeq
But the two
set of boundary conditions in eqs.(\ref{fbcu}) and (\ref{fbcp}) must be
consistent with each other. This implies $\eta_3=\eta_4$.

Let us now perform the conformal transformation given in eq.(\ref{ctz}) on
the
previous open string  boundary conditions in order to pass to the closed
string channel.
Since the right and left fermionic coordinates $\psi_-$ and ${\psi}_+$ are
two-dimensional conformal fields with conformal weight $h=\frac{1}{2}$
with respect to the their variables $\zeta$ and $\bar\zeta $ respectively,
then under the conformal transformation
\beq
\label {ctzz}
\zeta\rightarrow f(\zeta)=-i\zeta~~~~~{\rm and}~~~~~
\bar\zeta \rightarrow \bar f(\bar\zeta)=i\bar\zeta
\eeq
they transform as
\beq
\psi_-(\zeta)\rightarrow\psi'_-(\zeta)=
\left( \frac{\partial f(\zeta)}{\partial
\zeta} \right)^{1/2}
\psi_- (\zeta')
=(-i)^{\frac{1}{2}} \psi_- (f(\zeta))
\label{ii}
\eeq
and
\beq
\label {iii}
\psi_+(\bar\zeta)\rightarrow\psi'_+(\bar\zeta)
=\left(\frac{\partial f(\bar\zeta)}{\partial \bar\zeta}\right)^{1/2}
\psi_+ (\bar\zeta)
=(i)^{\frac{1}{2}} \psi_+ (\bar f(\bar\zeta))
\eeq
This implies that, performing the previous transformation on
eq.(\ref{fbcuS}),
we get  a relative factor $i$ between the right and left modes. More
specifically from the boundary conditions given
in eqs.(\ref{fbcuS}) and (\ref{fbcp})
after the conformal rescaling in eq.(\ref{res}) we get
\beq
\label{fbcucl}
\left\{
\begin{array}{l}
\psi^\mu_-(0,\sigma)=i\eta_1S^{\mu}\,\,\,_{\nu}{\psi^\nu_+}(0,\sigma)\\
\psi^\mu_-({\hat{T}}, \sigma)=i\eta_2S^{\mu}_{\,\,\, \nu}
{\psi^\nu_+}({\hat{T}}, \sigma)
\end{array}
\right.
\eeq
and
\beq
\label{fbcpcl}
\left\{
\begin{array}{l}
\psi^\mu_-(0,\tau )=\eta_3\psi^\mu_-(\pi,\tau)\\
{\psi^\mu_+}(0,\tau)=\eta_3{\psi^\mu_+}(\pi,\tau)
\end{array}
\right.
\eeq
where we have explicitly put $\eta_4=\eta_3$.

If we now compare the  usual boundary conditions
for the closed superstring theory
given in eqs. (\ref{ucbc}) and (\ref{rns2})
with eq. (\ref{fbcpcl})  we see that,
as a consequence of the identity between  $\eta_3$ and $\eta_4,$
the fermionic boundary state has only the R-R  and the NS-NS sectors.

As for the bosonic coordinate the boundary state for the fermionic
coordinate
at $\tau=0$ is defined, from the first equation in (\ref{fbcucl}), as the
state
that satisfies the equation:
\beq
\label {psib}
(\psi^\mu_-(0,\sigma)-i\eta S^{\mu}_{\,\,\, \nu} {\psi^\nu_+}(0,\sigma))
|B_{\psi}, \eta \rangle  =0
\eeq
where $\eta = \pm 1$.
Using the mode expansion given in eqs. (\ref{modpsil2}) and (\ref{modpsir})
we get the overlap conditions for the fermionic boundary state
\beq
\label{feroverl}
\left(\psi^\mu_t-i \eta S^{\mu}_{\,\,\, \nu} {\widetilde\psi^\nu_{-t}}
\right)|B_{\psi} , \eta \rangle  =0
\eeq
where the index $t$ is  integer [half-integer] in the R-R [NS-NS] sector.

In the case of the NS-NS-sector the determination of the  fermionic boundary
state satisfying eq.(\ref{feroverl}) is straightforward and leads to the
following expression:
\beq
\label{bnsns}
|B_{\psi} , \eta \rangle  = -i \prod_{r=1/2}^\infty 
\left(e^{i\eta\psi_{-r}\cdot S\cdot
\widetilde \psi_{-r}} \right) |0\rangle
\eeq
In the R-R sector the boundary state has the same form as in the NS-NS
sector
for what the non-zero modes is concerned, but with integer instead of
half-integer modes. We get therefore
~\footnote{The unusual phases introduced in
Eqs. (\ref{bnsns}) and (\ref{brr}) will turn out to be convenient
to study the couplings of the massless closed string states
with a D-brane and to find the correspondence with the
classical D-brane solutions obtained from supergravity. Note that
these phases are instead irrelevant when one computes the interactions
between two D-branes.}
\beq
\label{brr}
|B_{\psi} , \eta \rangle  =-\prod_{m=1}^\infty e^{i\eta\psi_{-m}\cdot
S\cdot\widetilde
\psi_{-m}} |B_{\psi} , \eta \rangle ^{(0)}
\eeq
where the zero mode contribution $|B_{\psi}, \eta \rangle ^{(0)}$ must
satisfy the
condition
\beq
\label{over0}
\left(\psi^\mu_0-i\eta S^{\mu}_{\,\,\, \nu}
{\widetilde\psi^\nu_{0}}\right) |B_{\psi}, \eta \rangle ^{(0)} = 0
\eeq
The previous equation is satisfied by the state
\beq
\label{solover}
|B_\psi , \eta\rangle ^{(0)} = {\cal M}_{AB}|A\rangle |\widetilde B\rangle
\eeq
where
\beq
\label{soloverM}
{\cal M}_{AB}=\left(C\Gamma^0...\Gamma^p\frac{1+i\eta\Gamma^{11}}{1+i\eta}
\right)_{AB}
\eeq
where $C$ is the charge conjugation matrix and $\Gamma^\mu$ are the Dirac
$\Gamma$ matrices in the $10$-dimensional space (see Ref.~\cite{cpb} for some
detail about the derivation of eqs.(\ref{solover}) and (\ref{soloverM})).

The overlap conditions for the conjugate boundary state can be obtained
from eq.(\ref{feroverl}) by taking the adjoint of this equation
namely
\beq
\label{conj4}
\langle B_\psi,\eta|\left(\psi^\mu_{-t}+i \eta S^{\mu}_{\,\,\, \nu}
{\widetilde\psi^\nu_{t}}
\right)=0
\eeq
and are solved by
\beq
\label{conj5}
\langle B_\psi,\eta|_{\rm NS}=i \langle0|\prod_{r=1/2}^\infty \left(
e^{i\eta\psi_{r}\cdot S\cdot
\widetilde \psi_{r}} \right)
\eeq
in the NS-NS sector and by
\beq
\label{conj6}
\langle B_\psi,\eta|_{\rm R}=-\langle B_\psi,\eta|^{(0)}_R
\prod_{m=1}^\infty
e^{i\eta\psi_{m}\cdot S\cdot
\widetilde \psi_{m}}
\eeq
in the R-R sector, where the zero mode is given by
\beq
\label{conj7}
\langle B_\psi,\eta|^{(0)}_{\rm R}=\langle A|\langle\widetilde B| ~{\cal
N}_{AB}
\eeq
with
\beq
\label{conj8}
 {\cal
N}_{AB}=(-1)^p\left(C\Gamma^0...\Gamma^p\frac{1+i\eta\Gamma^{11}}{1-i\eta}
\right)_{AB}
\eeq
Notice that the previous overlap conditions for the conjugate boundary state 
differ from those given in Ref.~\cite{BILLO} by the exchange 
$\eta\rightarrow-\eta.$
Our choice of $\eta$ corresponds to keep also in the closed channel
the same $\eta_{1},\eta_{2}$ appearing  in the open
string boundary conditions (see eq.(\ref{fbcuS})).

As in the bosonic string we must also in this case introduce a boundary
state for the reparametrization ghosts $b, c$. Moreover we must also add the 
boundary state for the superghosts $\beta, \gamma$.
The
complete boundary state for both the NS-NS and R-R sectors is given by:
\beq
|B, \eta \rangle _{R,NS} = \frac{T_p}{2} |B_{mat}, \eta \rangle  | B_{g},
\eta \rangle
\label{bounda3}
\eeq
where
\beq
|B_{mat} \rangle  = |B_X \rangle  |B_{\psi}, \eta \rangle
\hspace{1cm};\hspace{1cm}
|B_{g}\rangle  = |B_{gh} \rangle  | B_{sgh}, \eta \rangle
\label{bounda4}
\eeq
The matter part of the boundary state consists of the boundary state for the
bosonic coordinate $X$ given in eq.(\ref{b1}) without the normalization factor
$N_p$ and of the one for the fermionic
coordinate $\psi$ given in eq.(\ref{bnsns}) for the NS-NS sector and in
eq.(\ref{brr}) for the R-R sector. The ghost part $|B_g \rangle $ contains
the
boundary state corresponding to the ghosts $(b,c)$ given in eq.(\ref{bgh})
and the one corresponding to the superghosts $(\beta, \gamma )$
that we now want to determine. 

It is not difficult to check that the
identifications (\ref{over1}) and (\ref{feroverl})
imply that $\ket{B_{\rm mat}, \eta}$ is annihilated by the following
linear combinations of left and right generators of
the super Virasoro algebra
\begin{equation}
\left( { L}_n^{\rm mat} - {\tilde{{ L}}}_{-n}^{\rm mat} \right)
\ket{B_{\rm mat}, \eta} =0 ~~~,~~~
\left( { G}_m^{\rm mat} + \ii \eta  {\tilde{G}}_{-m}^{\rm mat} \right)
\ket{B_{\rm mat}, \eta} =0~~.
\label{viraco}
\end{equation}
The boundary state $\ket{B, \eta}$ must be BRST invariant,
that is
\begin{equation}
\label{bs1}
\left({ Q} + { \tilde Q}\right)\ket{B,\eta} = 0~~,
\end{equation}
where the BRST charge introduced in eq.(\ref{sbrchar}) is equal to
\beq
Q = \oint \frac{dz}{2 \pi i} \left[ c (z) \left( T^{mat} (z) + \frac{1}{2}
T^{g} (z) \right) - \gamma (z) \left( G^{mat}  (z) + \frac{1}{2} G^{g} (z)
\right) \right]~.
\label{BRST34}
\eeq
If we write the previous expression in the following form
\beq
Q = Q^{(1)} + Q^{(2)}~,
\label{12}
\eeq
where
\beq
Q^{(1)} = \sum_n c_n L_{-n}  - \sum_t \gamma_t G_{-t}
\label{q13}
\eeq
and
\beq
2 Q^{(2)} =\sum_{n=2}^{\infty} L_{-n}^{g} c_n + \sum_{n=-1}^{\infty} c_{-n} 
L_{n}^{g} - \sum_{t > 3/2} G_{-t}^{g} \gamma_t - \sum_{t > - 1/2} \gamma_{-t}
G_{t}^{g}~,
\label{q23}
\eeq 
it is easy to show that eqs.(\ref{viraco}) and (\ref{bs1}) imply
\begin{eqnarray}
\label{bsghost}
\left(c_{n} + {\tilde{c}}_{-n}\right) \ket{B_{\rm gh} } = 0~~~~, & &
\left(b_{n} - {\tilde{b}}_{-n}\right) \ket{B_{\rm gh}} = 0~~,
\nonumber\\
\left(\gamma_{t} +\ii\eta {\tilde{\gamma}}_{-t}\right)
\ket{B_{\rm sgh} ,\eta} = 0~~~~,
& &
\left(\beta_{t} +\ii \eta {\tilde{\beta}}_{-t}\right)
\ket{B_{\rm sgh} ,\eta} =
0~~.
\end{eqnarray}
Those equations imply that  the relations in eqs.(\ref{viraco}) must be 
supplemented by the analogous ones in the ghost sector, namely
\begin{equation}
\left( {L}_{n}^{\rm g} - {\tilde{L}}_{-n}^{\rm g} \right)
\ket{B_{\rm g}, \eta} =0
~~~,~~~
\left({G}_{m}^{\rm g} + \ii \eta  {\tilde{{ G}}}_{-m}^{\rm g}
\right) \ket{B_{\rm g}, \eta} =0  ~~.
\label{lngh}
\end{equation}

The overlap equations involving the ghost fields $b$ and $c$ can be solved and
one obtains the boundary state for the $bc$ system given in eq.(\ref{bgh}).
On the other hand the overlap equations for the superghosts determine the
superghost boundary state to be:
\begin{equation}
\label{bs8}
\ket{B_\sgh,\eta}_{\rm NS} =
\exp\biggl[\ii\eta\sum_{r=1/2}^\infty(\gamma_{-r}
\tilde\beta_{-r} - \beta_{-r}
  \tilde\gamma_{-r})\biggr]\,
  \ket{P=-1}\,\ket{\tilde P=-1}~,
\end{equation}
in the NS sector in the picture $(-1,-1)$ and
\begin{equation}
\label{bs10}
\ket{B_\sgh,\eta}_\R =
\exp\biggl[ \ii\eta\sum_{m=1}^\infty(\gamma_{-m}
\tilde\beta_{-m} - \beta_{-m}
\tilde\gamma_{-m})\biggr]\,
 \ket{B_\sgh,\eta}_\R^{(0)}~~,
\end{equation}
in the R sector in the $(-1/2,-3/2)$ picture.
The superscript $^{(0)}$ denotes the zero-mode contribution
that, if $\ket{P=-{1/ 2}}\,\ket{\tilde P=-{3/ 2}}$
denotes the superghost vacuum  that is annihilated by
$\beta_0$ and $\tilde \gamma_0$, is given by~\cite{YOST}
\beq
\label{bsrsg0}
\ket{B_\sgh,\eta}_\R^{(0)} =
\exp\left[\ii\eta\gamma_0\tilde\beta_0\right]\,
  \ket{P=-{1/ 2}}\,\ket{\tilde P=-{3/ 2}}~~.
\eeq
The conjugate boundary state for the superghost is equal to
\begin{equation}
  \label{bs27}
  {}_\NS\bra{B_\sgh,\eta} = \bra{P=-1}\,\bra{\tilde P=-1}\,
  \exp\biggl[-\ii\eta\sum_{m=1/2}^\infty(
  \beta_m \tilde\gamma_m - \gamma_m
  \tilde\beta_m)\biggr]
\end{equation}
in the NS sector and 
\[
{}_\R\bra{B_\sgh,\eta}=\bra{P=-{3/ 2}}\,\bra{\tilde P=-{1/2}}\,
  \exp\left[- \ii\eta\beta_0\tilde\gamma_0\right] \times
\]
\begin{equation}
\label{bs29}
\times
  \exp\biggl[- \ii\eta\sum_{m=1}^\infty(
  \beta_m \tilde\gamma_m -\gamma_m \tilde\beta_m )\biggr]
\end{equation}
in the R-R sector.
\par
We would like to stress that the boundary states $\ket{B}_{\NS,\R}$
are written in a definite picture $(P,\tilde P)$
of the superghost system, where $P$ is given in eq.(\ref{pict})
and $\tilde P= -2 -P$ in order to soak up the anomaly in
the superghost number. In particular we have chosen $P=-1$ in the NS
sector and $P=-1/2$ in the R sector, even if other choices would have been
in principle possible \cite{YOST}. Since $P$ is half-integer in the R sector,
the boundary state $\ket{B}_{\R}$ has always $P\not=\tilde P$, and thus 
it can couple only to R-R states in the asymmetric picture
$(P,\tilde P)$. However, as we have seen in section~\ref{sec:third} 
the massless R-R states in the $(-1/2,-3/2)$ picture 
contain a part that is proportional to the R-R {\it potentials}~\cite{sagnotti},
as opposed to the standard massless R-R states in the
symmetric picture $(-1/2,-1/2)$ that are always proportional to the R-R field
strengths. 

The boundary state in eq.(\ref{bounda3}) depends on the two values of
$\eta = \pm 1$.
Actually, as we will now show, we have to take a combination of the two
values
of $\eta$ corresponding to the GSO projection. Let us start with the NS
sector.
In the NS-NS sector the GSO projected boundary state is
\begin{equation}
\label{bs22a}
\ket{B}_\NS  \equiv  {1 +(-1)^{F+G}\over 2}\,\, {1 +(-1)^{\widetilde
F+\widetilde
G}\over 2}\, \ket{B,+}_\NS ~~,
\end{equation}
where $F$ and $G$ are the fermion and superghost number operators
\begin{equation}
{F} = {\sum_{m=1/2}^{\infty}\psi_{-m} \cdot \psi_m}-1
~~,~~
{G} = {- \sum_{m=1/2}^{\infty}
\left( \gamma_{-m}  \beta_m + \beta_{-m} \gamma_m
\right)} ~~.
\label{fergh}
\end{equation}
Their action on the boundary state corresponding to the fermionic coordinate
$\psi$ and to the superghosts can easily be computed and one gets:
\beq
(-1 )^{F} |B_{\psi} , \eta\rangle  = - |B_{\psi} , -\eta\rangle
~~~~;~~~~
(-1 )^{{\widetilde{F}}} |B_{\psi} , \eta\rangle  = - |B_{\psi}
, -\eta\rangle
\label{acti3}
\eeq
\beq
(-1 )^{G} |B_{sgh} , \eta\rangle  =  |B_{sgh} , -\eta\rangle
~~~~; ~~~~
(-1 )^{{\widetilde{G}}} |B_{sgh} , \eta\rangle  =  |B_{sgh} , -\eta\rangle
\label{acti4}
\eeq
Using the previous expressions after some simple algebra we get 
\begin{equation}
\label{bs22ab}
\ket{B}_\NS
= {1\over 2} \Big( \ket{B,+}_\NS - \ket{B,-}_\NS \Big)
\end{equation}
Passing to the R-R sector the GSO projected boundary state is
\begin{equation}
\label{bs22ba}
\ket{B}_\R  \equiv  {1 +(-1)^{p} (-1)^{F+G}\over 2}\,\,
{1 - (-1)^{\widetilde F+\widetilde G}\over 2}\, \ket{B,+}_\R ~~.
\end{equation}
where $p$ is even for Type IIA and odd for Type IIB, and
\begin{equation}
(-1)^{F} = \psi_{11}(-1)^{\sum\limits_{m=1}^{\infty}\psi_{-m}
\cdot \psi_m}
~~,~~
{G} = - \gamma_0 \beta_0 - \sum\limits_{m=1}^{\infty}
\left[ \gamma_{-m} \beta_m + \beta_{-m}\gamma_m \right]~~.
\label{fermion}
\end{equation}
From the previous expressions it is easy to see after some calculation that
the action of the fermion number operators is given by:
\beq
(-1)^{F} |B_{\psi}, \eta \rangle  = (-1)^p |B_{\psi}, - \eta \rangle
\hspace{.5cm};
\hspace{.5cm}
(-1)^{{\widetilde{F}}} |B_{\psi}, \eta \rangle  =  |B_{\psi}, - \eta \rangle
\label{}
\eeq
and
\beq
(-1)^{G} |B_{sgh}, \eta \rangle  =   |B_{sgh}, - \eta \rangle
\hspace{.5cm};\hspace{.5cm}
(-1)^{{\widetilde{G}}} |B_{sgh}, \eta \rangle  = -  |B_{sgh}, - \eta \rangle
\label{fernnu3}
\eeq
Using the previous expressions after some straightforward manipulations, one
gets
\beq
\label{bs22bb}
\ket{B}_\R  =
    {1\over 2} \Big( \ket{B,+}_\R + \ket{B,-}_\R\Big)~~.
\end{equation}

\section{Classical Solutions From Boundary State}
\label{sec:eighth}

In this section we want to connect the boundary state introduced in the
previous sections to the Dirichlet branes intended as electric R-R charged 
$p$-brane solutions of the low-energy string effective action. 
In particular we  will show that the
large distance behaviour of the graviton, dilaton and R-R $p+1$-form fields 
that one obtains from the boundary state exactly agrees with that obtained
from the classical solution in sect.~\ref{sec:fifth}.  

The long distance behaviour of the classical massless fields
generated by a {\dpb} can be determined by computing the projection of
the boundary state along the various fields after having inserted a closed
string propagator. This amounts to compute the following matrix element
\beq
\label{filflu}
\langle P_x|D|B\rangle
\eeq
where $P_x$ runs over all the projectors of the closed superstring massless
sector listed in Ref.~\cite{antone}, $D$ is the propagator in eq.(\ref{propg})
if we perform the calculation in the bosonic string or is given by
\beq
D = \frac{\alpha'}{4 \pi} \int \frac{d^2 z}{|z|^2} z^{L_0 - a} {\bar{z}}^{
{\tilde{L}}_0 - a}
\label{propg5}
\eeq
if we more correctly perform the calculation in superstring,
where the constant $a =1/2$ in the NS-NS sector and $a=0$ in the R-R sector.

Let us start by computing  the expression for the generic NS-NS massless 
field which is given by
\beq
J^{\mu \nu} \equiv {}_{-1}
\langle  {\widetilde{0}}| {}_{-1}\langle 0| \psi^{\nu}_{1/2}~ 
{\widetilde{\psi}}^{\mu}_{1/2}  |D|
B\rangle _{NS}= - \frac{T_p}{2k^2_\perp} V_{p+1}S^{\nu\mu}
\label{masscam}
\eeq
This equation is exactly the same of the one that one gets in the bosonic string
(except that in this case $d=10$ and not $d=26$)
if we use the propagator in eq.(\ref{propg}), the boundary state in 
eq.(\ref{b1}) and the bosonic massless closed string state $\langle  
{\widetilde{0}}| \langle 0| \alpha^{\nu}_{1}~ 
{\widetilde{\alpha}}^{\mu}_{1}$. Because of this we keep the value of the 
space-time dimension $d$ arbitrary in such a way that our calculation is 
valid in both cases. 
Specifing the different polarizations corresponding to the various fields
(see Refs.~\cite{cpb,antone} for details) we get
\beq
\label{dilatone}
\delta \phi=\frac{1}{\sqrt{d - 2}}\,\left(\eta^{\mu\nu} -
k^{\mu} \ell^{\nu} - k^{\nu}\ell^{\mu}\right)
J_{\mu\nu}=
 \frac{d-2p-4}{2{\sqrt {2(d-2)}}}\,\mu_p\,
\frac{V_{p+1}}{k_{\bot}^{2}}
\eeq
for the dilaton,
\[
\delta h_{\mu \nu}(k) =\frac{1}{2}\Big(J_{\mu\nu}+J_{\nu\mu}\Big)
-\frac{\delta\phi}{\sqrt{d-2}}\, \,\eta_{\mu\nu}=
\]
\beq
=\sqrt 2 \mu_p \frac{ V_{p+1}}{ k_{\bot}^2}\,
{\rm diag} \left(- A, A \dots A, B \dots B \right)~~,
\label{gravitone}
\eeq
for the graviton,
where $A$ and $B$ are given in eq. (\ref{expo});
and
\beq
\delta B_{\mu\nu}(k) = \frac{1}{\sqrt 2}\Big( J_{\mu\nu} - J_{\nu\mu}
\Big)=0
\label{ant1}
\eeq
for the antisymmetric tensor. In the R-R sector we get instead
\beq
\delta C_{01\dots p}(k) \equiv \bra{P^{(C)}_{01\cdots p}} \,D\,\ket{B}_{\rm
R} =\mp\,{\mu_p}
\frac{V_{p+1}}{k_{\bot}^2}
\label{ra1}~~.
\eeq
Expressing the previous fields in configuration space
using  the following Fourier transform valid for $p < d-3$
\begin{equation}
\int d^{(p+1)}x\,d^{(d-p-1)} x \frac{{\rm e}^{i k_{\bot} \cdot x_{\bot}}}{
(d-p-3)\,
r^{d-p-3}\,\Omega_{d-p-2}}= \frac{V_{p+1}}{k^2_\perp}~,
\label{ft}
\end{equation}
remembering the expression $Q_p$ defined in eq.(\ref{Qp})
and  rescaling the fields according to
\beq
{\varphi} = { \sqrt{2}\kappa} {{\phi}} ~~~~,~~~~
{\tilde h}_{\mu \nu} = {2 \kappa} {{h}}_{\mu \nu} ~~~~,~~~~
{{ {\cal{C}}}}_{01...p} = {\sqrt{2}}{\kappa}
{C}_{01...p}~~, \label{newfi}
\eeq
we get the following large distance behaviour
\beq
\label{dilatone2}
\delta\varphi(r)=   \frac{d-2p-4}{{2\sqrt{2(d-2)}}}\,
\frac{Q_{p}}{r^{d-p-3}}
\eeq
for the dilaton,
\beq
\delta {\tilde h}_{\mu \nu}(r) = 2\frac{Q_{p}}{ r^{d-p-3}}\,
{\rm diag} \left(- A,  \dots A, B \dots B \right)~~,
\label{gravitone2}
\eeq
for the graviton and
\beq
\label{RR2}
{\delta{{\cal C}}}_{01...p} = \mp\frac{Q_{p}}{r^{d-p-3}}
\eeq
for the R-R form potential.

The previous equations reproduce exactly the behavior for $r\to \infty$ of 
the metric in eq.(\ref{metri}) and of the R-R potential given in 
eq.(\ref{dil}). In fact at large distance their fluctuations around the 
background values are exactly equal to ${\delta{\tilde h}}_{\mu\nu}$
and ${\delta{{{\cal C}}}}_{01...p}$. In the case of the dilaton, in
order to find agreement between the boundary state and the classical solution,
we have to take $d=10$. This strongly suggests that, as expected, the 
calculation has to be performed in superstring. 
As a matter of fact, a comparison between the $p$-brane solution
of the classical eqs. of motion that follow from the action in (\ref{action}) 
and a string calculation, does make sense only in the
superstring case where the graviton, dilaton and Kalb-Ramond field
come from the NS-NS sector and the antisymmetric
gauge potentials like ${\cal{C}}_{\mu_1...\mu_{n}}$
from the R-R sector.
Nonetheless, the bosonic case we have also considered in this section already
tells us what are the distinctive
features of the boundary state and how the
long-distance behavior of the massless fields is encoded in it.

\section{Interaction Between a $p$ and a $p'$  Brane}
\label{sec:ninth}

In this section we study the static interaction between a D$p$-brane
located at $y_1$, and a D$p'$-brane located at
$y_2$, with $NN\equiv{\rm min} \{p,p'\} +1$ directions common to the brane
world-volumes, $DD\equiv {\rm  min}\{d-p-1,d-p'-1\}$
directions transverse to both,
and $\nu = (d -NN -DD)$ directions of mixed type.
We will not consider instantonic D-branes, hence also $NN\geq 1$.
The two D-branes simply interact via tree-level exchange of
closed strings whose propagator is
\beq
D= {\alpha'\over 4\pi} \int{d^2 z\over |z|^2} z^{L_0 }\,
\bar z^{\widetilde L_0 }~~,
\label{propp}
\end{equation}
so that  indicating with $\ket{B_1}$ and $\ket{B_2}$ the boundary states
describing the two D-branes the static amplitude
is given by
\begin{equation}
  \label{bs32}
 A = \bra{B_1}~D~\ket{B_2}=\frac{T_p\,T_{p'}}{4}
\frac{\alpha'}{4\pi}\int_{|z|<1}
\frac{d^2z}{|z|^2}{\cal A}\,{\cal A}^{(0)}~~,
\end{equation}
where we have indicated with
${\cal A}$ and  ${\cal A}^{(0)}$ respectively
the non zero mode and the zero mode contribution in which the previous
amplitude can be factorized. We do not have any intercept as
we had in eq.(\ref{propg}) for the bosonic string because we assume that both
$L_0$ and ${\widetilde{L}}_0$ contain the ghost degrees of freedom. The
details of the computation of the quantity in eq.(\ref{bs32}) can be found
in Ref.~\cite{BILLO}. Here we just give the results of the various terms 
starting
from the non-zero modes. In the NS-NS sector after the GSO projection we get
\beq
\label{ansnz}
{\cal A}_{\rm{NS-NS}}= \frac{1}{2}
\left[ \left({f_3\over f_1}\right)^{8 -\nu}
  \left({f_4\over f_2}\right)^{\nu}
- \left({f_4\over f_1}\right)^{8 -\nu}
  \left({f_3\over f_2}\right)^{\nu}\right]~~,
\eeq
In the R-R sector instead before the GSO projection we get
\beq
\label{arrnz}
{\cal A}_{\rm{R-R}}(\eta_1,\eta_2) =
 \left[ 2^{\nu-4}
\left(\frac{f_2}{f_1}\right)^{8-2\nu}
\delta_{\eta_1\eta_2,1}+\delta_{\eta_1\eta_2,-1}\right]~,
\eeq
where the functions $f_i$ are equal to 
\beq
\label{f12}
f_1\equiv q^{{1\over 12}} \prod_{n=1}^\infty (1 - q^{2n})~~~~;~~~~
f_2\equiv \sqrt{2}q^{{1\over 12}} \prod_{n=1}^\infty (1 + q^{2n})~~;
\eeq
\beq
\label{f34}
f_3\equiv q^{-{1\over 24}} \prod_{n=1}^\infty (1 + q^{2n -1}) ~~~~;~~~~
f_4\equiv q^{-{1\over 24}} \prod_{n=1}^\infty (1 - q^{2n -1}) ~~.
\eeq
They transform as follows under the modular transformation 
$t\rightarrow 1/t$ ($q=e^{- \pi t}$):
\beq
\label{mtfi}
f_1(e^{-\frac{\pi}{t}})=\sqrt{t}f_1(e^{-\pi t})~~;~~
f_2(e^{-\frac{\pi} {t}})=f_4(e^{-\pi t})~~;
f_3(e^{-\pi t})=f_3(e^{-\frac{\pi}{ t}})~.
\eeq
The zero modes contribution in the NS-NS sector comes only from the bosonic 
coordinate and  can be obtained by following the same procedure outlined in
eqs.(\ref{zp})-(\ref{zpp}) for the bosonic string. Also in this way one gets
eq.(\ref{zpp}). If we insert the contributions in eqs.(\ref{ansnz}) and 
(\ref{zpp}) in eq.(\ref{bs32}) we get the total NS-NS contribution to the 
interaction between a D$p$ and a D$p'$ brane
\begin{eqnarray}
\label{ansns}
A_{\rm{NS-NS}}
&=& V_{NN}\, (8\pi^2\alpha')^{-{NN\over 2}}
\int_0^\infty
{dt} \left(1\over t \right)^{DD\over
2}\,{\rm e}^{- y^2 /(2\alpha' \pi t)}\,\nonumber\\
& & \times \frac{1}{2}\left[ \left({f_3\over f_1}\right)^{8 -\nu}
  \left({f_4\over f_2}\right)^{\nu}
  - \left({f_4\over f_1}\right)^{8 -\nu}
  \left({f_3\over f_2}\right)^{\nu}\right]~~,
\end{eqnarray}
where $V_{NN}$ is the common world-volume of the two D-branes,
$|y|$ is the transverse distance between them.

It is interesting to notice that
the two terms in the square brackets of \eq{ansns} come respectively
from the NS-NS  and the NS-NS$(-1)^{(F+G)}$ sectors of the
exchanged closed string, which, under the
transformation $t = {1}/{\tau}$, are mapped into the
NS and R sectors of the open string suspended between
the branes. Notice that $A_{\rm{NS-NS}}=0$ if
$\nu= 4$.
\par
The evaluation of the zero mode contribution
in the R-R sector requires more care due to the presence of
zero modes both in the
fermionic matter fields and the bosonic superghosts.
Inserting eq. (\ref{arrnz}) into eq. (\ref{bs32}) we can write
the total R-R contribution as
\[
A_{\rm{R-R}}(\eta_1,\eta_2) =
 V_{NN}\, (8\pi^2\alpha')^{-{NN\over 2}}
2^{-{\nu\over 2}}\int_0^\infty
{dt}
\left( 1\over t \right)^{DD\over 2}\,
{\rm e}^{- y^2 /(2\pi \alpha' t)}
\]
\beq
\times \left[ 2^{\nu-4}
\left(\frac{f_2}{f_1}\right)^{8-2\nu}
\delta_{\eta_1\eta_2,+1}+\delta_{\eta_1\eta_2,-1}\right]
~{}_\R^{(0)}\!\langle B^1,\eta_1 |
B^2,\eta_2\rangle_\R^{(0)}~~,
\label{arr1}
\eeq
where
\beq
\ket{B,\eta}_\R^{(0)} =
\ket{B_\psi,\eta}_\R^{(0)}~\ket{B_\sgh,\eta}_\R^{(0)}~~.
\label{bsr00}
\eeq
Note that
in \eq{arr1} it is essential {\it not} to separate the matter
and the superghost zero-modes.
In fact, a na{\"{\i}}ve evaluation of
${}_\R^{(0)}\!\langle B^1,\eta_1 |
B^2,\eta_2\rangle_\R^{(0)}$ would lead to a divergent or
ill defined result:
after expanding the exponentials in
${}_\R^{(0)}\!\langle B^1_\sgh,\eta_1 |
B^2_\sgh,\eta_2\rangle_\R^{(0)}$, all the infinite terms
with any superghost number contribute, and yield the divergent sum
$1+1+1+...$ if $\eta_1\eta_2=-1$, or the alternating sum $1-1+1-...$ if
$\eta_1\eta_2=1$. This problem has already
been addressed in Ref.~\cite{YOST} and solved by introducing
a regularization scheme for the pure Neumann case ($NN=10$).
This method has been extended to the most general case with D-branes in
Ref.~\cite{BILLO}. Here, we give the final result for the fermionic zero mode
part of the R-R sector:
\beq
{}_\R^{(0)}\!\langle B^1,\eta_1 |
B^2,\eta_2\rangle_\R^{(0)}  =
-16 \,\delta_{\nu,0}\,\delta_{\eta_1\eta_2,1} + 16 \,
\delta_{\nu,8}\,\delta_{\eta_1\eta_2,-1}~~.
\label{result3}
\eeq

We can now write the final expression for the R-R amplitude. Inserting
\eq{result3} into \eq{arr1}, after performing the GSO projection we get
\[
 A_{\rm{R-R}}
= V_{NN}  (8\pi^2\alpha')^{-{NN\over 2}} \cdot
\]
\beq
\int_0^\infty {dt}
\left(1\over t \right)^{DD\over 2}\,
{\rm e}^{-  y^2 /(2 \pi \alpha' t)}
~\frac{1}{2} \left[ -\left(\frac{f_2}{f_1}\right)^{8}\,
\delta_{\nu,0}+\delta_{\nu,8}\right]~~.
\label{arr3}
\eeq
The $\nu=0$ and $\nu=8$ terms in \eq{arr3} come respectively from
the R-R and the R-R$(-1)^{(F+G)}$ sectors of the exchanged
closed string, which, under the transformation $t\to 1/t$,
are mapped into the ${\rm NS}(-1)^{(F+G)}$ and ${\rm R}(-1)^{(F+G)}$
sectors of the open string suspended between the branes.
Due to the ``abstruse identity'', the total D-brane amplitude
\beq
A=A_{\rm{NS-NS}}+A_{\rm{R-R}}
\label{total}
\eeq
vanishes if $\nu=0,4,8$; these are precisely the configurations of
two D-branes which break half of the supersymmetries of the Type II
theory and satisfy the BPS no-force condition.

\acknowledgements 

We would like to thank M. Frau, T. Harmark, A. Lerda, R. Marotta, I. Pesando 
and R. Russo for many discussions on the subject of these lectures.

\vskip 1.3cm
\appendix{\Large {\bf{Appendix A}}}
\label{appe}
\vskip 0.5cm
\renewcommand{\theequation}{A.\arabic{equation}}
\setcounter{equation}{0}
\noindent
In this appendix we will describe the properties of bosonic and 
fermionic  $bc$ systems that enter in the covariant quantization of
string theories. Their dynamics is described by the action
\EQ
S[b,c] \sim  \int\! d^{2}z \,\,[b \bar{\partial} c +
\bar b {\partial} \bar c ]~~,
\label{2.1}
\EN
which implies the equations of motion
\EQ
\bar{\partial} b = 0 \hspace{2cm};\hspace{2cm} \bar{\partial} c=0~~,
\label{2.1a}
\EN
and their conjugate ones.
Thus, the fields $b$ and $c$ are functions only of $z$ and 
 they admit the following holomorphic expansions
\footnote{To avoid repetition, we write all definitions for
the holomorphic sector of the theory only; similar expressions hold
for the antiholomorphic sector.}
\EQ
b(z)= \sum b_{n} z^{-n-\lambda}\hspace{1.5cm};\hspace{1.5cm}
c(z)=\sum c_{n}z^{-n+\lambda-1}~~.
\label{2.2}
\EN
where the variable $n$ is integer for integer values of
the conformal dimension $\lambda$, while for  
half-integer values of $\lambda$ different spin structures are possible. 
In particular for periodic boundary conditions
(R-sector) $n$ is integer while for anti-periodic boundary conditions 
(NS-sector) $n$ is half-integer.
The oscillators in eq. (\ref{2.2}) satisfy the following hermiticity properties
\beq
\label{herm0}
c^\dagger_n=c_{-n}~~~~;~~~~b^\dagger_n=\epsilon b_{-n}~~,
\eeq
where $\epsilon =1$ for fermions and $\epsilon =-1$ for bosons.

The theory can be quantized by either requiring canonical commutation relations
that on the mode expansion in eq.(\ref{2.2}) read as
\EQ
[c_{n},b_{m} ]_{\epsilon} =\delta_{n+m,0}\hspace{1.5cm};\hspace{1.5cm}
[b_{n},b_{m}]_{\epsilon}=
[c_{n},c_{m}]_{\epsilon}=0~~,
\label{2.3}
\EN
where $[\,\, , \,\,]_{\epsilon}$ means commutator [anticommutator] for bosonic
[fermionic] fields  or by imposing the OPE
\EQ
c(z) b(w) = \frac{1}{z-w}~~~~~,~~~~~b(z) c(w) = \frac{\epsilon}{z-w}~~. 
\label{2.7a}
\EN
The energy-momentum tensor $T(z)$ and the
ghost number current $j(z)$ of the theory are given by 
\EQ
T(z)=:[-\lambda b \partial c +(1-\lambda)
\partial b  c]: \,\,\, = \sum_{n}L_n z^{-n-2}~~,
\label{2.6}
\EN
\beq
j(z)=-:b(z)c(z): \,\,\,= \epsilon\,\, : c(z) b(z) : = \sum_{n}j_n z^{-n-1}~~,
\label{2.6a}
\EN
where the normal ordering is explicitly given by
\EQ
: b_{n}c_{-n} :=\left\{ \begin{array}{ll}
                            b_{n} c_{-n} & \mbox{if $n< 1-\lambda$}\\
                           -\epsilon c_{-n}b_{n} & \mbox{if $ n \geq 1-\lambda$}
                          \end{array}
                  \right.~~.
\label{2.8}
\eeq
The Fourier coefficients $L_n$ and $j_n$
take the form
\bea
L_{n}&=&\oint\frac{dz}{2\pi i} T(z) z^{n+1}=
\sum_{m}(\lambda n -m) :b_{m}c_{n-m}:
\nonumber \\
j_{n}&=&\oint\frac{dz}{2\pi i} j(z) z^{n}=
-\sum_{m}:b_{m}c_{n-m}:~~,
\label{2.10}
\ena 
and as a consequence of eq. (\ref{herm0}) they satisfy the following 
hermiticity properties
\beq
\label{herm1}
L^\dagger_n=L_{-n}~~~~;~~~~j^\dagger_n=-j_{-n}~~.
\eeq
From the ghost number current we can obtain the ghost number
$j_0$ as
\EQ
j_{0} = \oint dz j(z) = - \sum_{n=-\infty}^{\infty} :b_{n} c_{-n}:~~,
\label{2.9}
\EN
By using the OPE in eq.(\ref{2.7a}) one gets
\beq
\label{A1}
T(z) b(w) = \frac{\lambda b(w)}{(z-w)^{2}} + 
\frac{\partial_w b(w)}{(z-w)}+..~;~ 
T(z) c(w) = \frac{(1-\lambda) c(w)}{(z-w)^{2}} + 
\frac{\partial_w c(w)}{(z-w)}+..~,
\eeq
that are consistent with the fact that $b$ and $c$
are conformal fields with conformal weights $\lambda$ and $1-\lambda$
respectively and
\beq
\label{A2}
j(z) b(w) = -\frac{b(w)}{(z-w)}+... ~~~~;~~~~ 
j(z) c(w) = \frac{c(w)}{(z-w)}+...~~, 
\eeq
that imply that $b$ and $c$ have ghost charge $-1$ and $1$ respectively.
Moreover one can see that
$T(z)$ and $j(z)$ satisfy the OPE
\bea
T(z) T(w) &=& \frac{c/2}{(z-w)^{4}} + 
\frac{2T(w)}{(z-w)^{2}} + \frac{\partial_{w}T(w)}{z-w} + \cdots
\nonumber \\
T(z) j(w) &=& \frac{{\cal Q}}{(z-w)^{3}} +
\frac{j(w)}{(z-w)^{2}} + \frac{\partial_{w}j(w)}{z-w} + \cdots
\nonumber \\
j(z) j(w) &=& \frac{\epsilon}{(z-w)^{2}} + \cdots~~,
\label{2.7}
\ena
where the "screening charge" ${\cal Q}$ and the $c$-number of the Virasoro algebra are
respectively given by
\beq
{\cal Q}= \epsilon(1-2\lambda )~~~~,~~~~c= \epsilon ( 1 - 3 {\cal Q}^2 )~~.
\label{Qc}
\eeq 
Using  eqs. (\ref{2.10}) the OPEs given in eq. (\ref{2.7}) imply
the commutation relations
\EQ
[L_{n},L_{m}] = (n-m) L_{n+m}+ \frac{1-3 {\cal Q}^{2}}{12} n(n^{2}-1)
\delta_{n+m,0}~~,
\label{2.7l} 
\EN
\EQ
[L_{n},j_{m}] = -m j_{n+m} +\frac{{\cal Q}}{2}n(n+1) \delta_{n+m,0}~~,
\label{2.7m} 
\EN
\EQ
[j_{n},j_{m}] = n \delta_{n+m,0}~~. 
\label{2.7b}
\EN
We observe that the ${\cal Q}$-dependent term in eq.(\ref{2.7m}) 
(or equivalently in the second OPE in eq.(\ref{2.7})), which
makes the current $j(w)$ not quite a good conformal
primary field, is a consequence of the anomaly appearing in the
conservation law of the ghost number current~\cite{FMS}
\beq
\label{anomalia}
\bar\partial j(z)=\frac{1}{8}{\cal Q}\sqrt{g}R^{(2)}~~,
\eeq 
where $g$ and $R^{(2)}$ are respectively, the determinant of the metric and
the scalar curvature of the two dimensional world-sheet $\Sigma$, on which the
theory is defined.


Comparing the two equations obtained from (\ref{2.7m}) for $n=-m=1$ and
$n=-m=-1$ and using the hermiticity properties in eq.(\ref{herm1}) 
we get that $j_0$ is neither hermitian nor
antihermitian, but satisfies the following property
\EQ
j_{0}+j_{0}^{\dagger}+{\cal Q}=0~~.
\label{2.9a}
\EN
Let us introduce the $q$-vacuum, $|q\!>$, defined by the relations
\bea
b_{n}|q\!>&=&0 \hspace{1cm} {\rm{if}} \hspace{1cm} n > \epsilon q -
\lambda~~,
\nonumber \\
c_{n}|q\!>&=&0 \hspace{1cm} {\rm{if}} \hspace{1cm}  n \geq -\epsilon q+\lambda
~~.\label{2.4}
\ena
Then, from eq.(\ref{2.10}) one can easily 
show that $|q\!>$ is an eigenstate of both $j_{0}$
and $L_{0}$ with eigenvalues given respectively by the following equations
\beq
j_0 |q\!> = q |q\!> ~~~~~,~~~~~ L_0 |q\!> =  \frac{1}{2} \epsilon q(q+{\cal Q})|q\!>
~~, 
\label{j0L0}
\eeq
and that, as a consequence of eq. (\ref{2.9a}), it is normalized as follows
\EQ
<\!q'|q\!> = \delta(q'+q+{\cal Q})
\label{2.9b}
\EN
We observe that the state $|q=0\!>$ is the only $SL(2,R)$ 
invariant vacuum since
it is the only one which is 
annihilated simultaneously by $L_{0},L_{1}$ and $L_{-1}$.

Because of eq.(\ref{2.9b}), in order not to get a vanishing result when
we compute correlation functions involving
$b$ and $c$, we must make sure that the total ghost number of the 
correlator be equal
$-{\cal Q}$. For instance the following correlation function
\EQ
<\!-q -{\cal Q}|c(z ) b(w) |q \!> = \left(\frac{z}{w} \right)^{\epsilon q}\frac{1}{z-w}~~,
\label{2.5}
\EN
is different from zero. The contraction given in eq.(\ref{2.7a}) can be 
obtained from the previous equation by choosing the 
$SL(2,R)$ invariant vacuum $|q=0\rangle$.

By using the mode expansion  in eq.(\ref{2.2}) and the anticommutation 
relations in eq. (\ref{2.3}), or equivalently the contraction 
in eq.(\ref{2.7a}) together with the Wick theorem,
one can very easily compute any correlation function of $b$ and $c$ fields
on the sphere. 

A fermionic $bc$ system can be bosonized in terms of a scalar field with a
background charge ${\cal Q}$ through the following relations
\EQ
b(z) =  :e^{-\varphi(z)}: \hspace{2cm} c(z) =
:e^{\varphi(z)}:~~,
\label{2.25}
\EN
while a bosonic $bc$ system can be "bosonized" in terms of a scalar field 
$\varphi$ with background charge ${\cal Q}$ and a fermionic $bc$ system with 
$\lambda =1$, that we call $\xi \eta$ system, through the following relations
\EQ
\beta (z) = \partial \xi(z) e^{-\phi(z)} \hspace{2cm} \gamma (z) =  e^{\phi(z)} 
\eta(z)~~.\label{bcboso}
\EN
where we have called $\beta, \gamma$ the bosonic $b,c$ fields.

Let us give some detail about this bosonization procedure.
The action of a scalar field with background charge ${\cal Q}$ is given by
\EQ
S[\varphi] \sim \int_{\Sigma} d^{2}z [ -\epsilon\bar{\partial}\varphi
\partial \varphi -\frac{1}{4}{\cal Q} \sqrt{g}R^{(2)}\varphi]
\label{2.18}
\EN
where $g$ and $R^{(2)}$ are respectively, the determinant of the metric and
the scalar curvature of the two dimensional world-sheet $\Sigma$, on which the
theory is defined. The equation of motion for this field is
\beq
\label{anomalia2}
\partial\bar\partial \varphi(z)=\frac{1}{8}\epsilon {\cal Q}\sqrt{g}R^{(2)}~~.
\eeq 
Notice that $\epsilon \partial\varphi(z)$ satisfies exactly the 
same equation as the anomalous current $j(z)$ of the previous system (see eq.
(\ref{anomalia})).      
This system is invariant under the conformal transformations generated by the
energy-momentum tensor
\EQ
T(z)=:\frac{1}{2}[ \epsilon (\partial \varphi)^{2}-{\cal Q}\partial^{2}\varphi]:(z)~~, 
\label{2.19a}
\EN
and under a $U(1)$ Kac-Moody algebra generated by the current
\EQ
j(z)= \epsilon \partial \varphi(z)~~. 
\label{2.19}
\EN
The theory can be quantized by requiring the standard OPE for a free scalar 
field, namely
\EQ
\varphi(z) \varphi(w) = \epsilon \log(z-w)~~. 
\label{2.26}
\EN
By using it one can easily check that $T(z)$ and $j(z)$
satisfy the following OPEs
\EQ
T(z) T(w) = \frac{c/2}{(z-w)^{4}}
+2\frac{T(w)}{(z-w)^{2}}+\frac{\partial_{w}T(w)}{z-w}+\cdots~~,
\label{2.20a}
\EN
\EQ
T(z) j(w) =
\frac{{\cal Q}}{(z-w)^{3}}+\frac{j(w)}{(z-w)^{2}}+\frac{\partial_{w}j(w)}{z-w}+
\cdots~~,
\label{2.20b}
\EN
\EQ
j(z) j(w) = \frac{\epsilon}{(z-w)^{2}}+\cdots~~,
\label{2.20c}
\EN
where the central charge $c$ of the Virasoro algebra is equal to
\beq
c = 1-3 \epsilon {\cal Q}^{2}~~.
\label{virc} 
\eeq
The presence of a third-order pole
in (\ref{2.20b}) is a signal of the fact that  $j(z)$ is not
really a good conformal field of weight 1, when there is a non-vanishing
background charge ${\cal Q}$.
Notice that eqs. (\ref{2.20a}) - (\ref{2.20c}) reproduce the OPE
given in eq. (\ref{2.7}) except for the value of the central charge $c$.  

The field $\varphi$ admits the following expansion
\EQ
\varphi(z)= x +N \log z + \sum_{n \neq
0}\frac{\alpha_{n}}{n}z^{-n}~~,
\label{2.21}
\EN
where the harmonic oscillators satisfy the usual commutation relations
\EQ
[\alpha_{n},\alpha_{m}] = n \epsilon \delta_{n+m,0} \hspace{2cm}
[x,N]=- \epsilon~~.
\label{2.22}
\EN
Using eq.(\ref{2.21}) in eqs.(\ref{2.19a}) and (\ref{2.19}), 
one can easily obtain
the oscillator expressions for the Virasoro generators $L_n$ and for the Fourier
components $j_n$ of the current $j(z)$, namely
\[
L_{n}=
\frac{1}{2}\sum_{m}:\alpha_{m}\alpha_{n-m}:
-\frac{1}{2}{\cal Q}(n+1)\alpha_{n}~~,
\]
\EQ
j_{n}=- \epsilon\alpha_{n}~~,
\label{2.24}
\EN
with $\alpha_{0}= - N$ and where the symbol $:\hspace{0.7cm}:$ is
the usual normal ordering of harmonic oscillators.
The OPEs in eqs.(\ref{2.20a}), (\ref{2.20b}) and (\ref{2.20c})
are then equivalent to the following commutation relations
\[
[L_n , L_m] = (n-m) L_{n+m} +
\frac{1-3 {\cal Q}^{2}}{12} n(n^{2}-1)
\delta_{n+m,0}~~,
\]
\EQ
[ L_n,j_m] = -m
j_{n+m} -\frac{{\cal Q}}{2}n(n+1) \delta_{n+m,0}~~;~~
[j_n , j_m] = n \delta_{n+m,0}~.
\label{2.24a}
\EN
As in the case of a $bc$ system the zero mode of the fermionic number current
is not hermitian, but satisfies the following hermiticity properties
\beq
j_0 + j_0^{\dagger} +{\cal Q} =0~~,
\label{hermi87}     
\eeq
which implies 
\EQ
<\!q\,|\,q'\!> = \delta (q+q'+{\cal Q})~~.
\label{2.24c}
\EN
where  $|q\!>$ and $|q'\!>$ are eigenstates of $N$ with
eigenvalues $q$ and $q'$ respectively

Due to the presence of the zero mode logarithmic term in eq.(\ref{2.21}), the
field $\varphi(z)$ does not transform properly under a conformal
transformation, whereas $\displaystyle{:e^{q \varphi(z)}:}$
behaves as a primary
conformal field of weight $\frac{1}{2} \epsilon q(q+{\cal Q})$. In addition it 
transforms as a field with charge $q$ under the ghost number current 
generated by $j(z)$. This can be checked by computing the following OPE
\EQ
T(z) :e^{q\varphi(w)}: = 
\frac{1}{2} \epsilon q(q+{\cal Q})\frac{:e^{q\varphi(w)}:}{(z-w)^{2}} +
\frac{\partial_{w}:e^{q\varphi(w)}:}{(z-w)} + \cdots~~,
\label{2.27}
\EN
\beq
j(z) :e^{q\varphi(w)}: = q \frac{:e^{q\varphi(w)}:}{z-w}~~.
\label{jexpo}
\eeq
Introducing the corresponding highest weight state according to
\EQ
|q\!>= \lim_{z \rightarrow 0} :e^{q\varphi(z)}:|0\!>~~,
\label{2.28}
\EN
it is easy to see that $|q\!>$ is an eigenstate of the ghost
number $j_0$ and of $L_0$ with eigenvalues given respectively by
\beq
L_0|q\!> = \frac{1}{2} \epsilon q(q + {\cal Q}) |q\!>~~~~,~~~~j_0 |q\!> = q |q\!>~~.
\label{L0j098}
\eeq 
If we consider the case $\epsilon=1$ and we takes ${\cal Q} =1 -2 \lambda$ we 
immediately see that the central charge in eq. (\ref{virc}) reproduces exactly
the one given in eq. (\ref{Qc}) and that
the OPEs in eqs.(\ref{2.7}) and in eqs.(\ref{2.20a}),
(\ref{2.20b}) and (\ref{2.20c}) are coincident. 
Moreover if we consider eqs. (\ref{2.27}) and  (\ref{jexpo}) for $\epsilon=1$
and ${\cal Q}=1-2\lambda$ and put $q=\pm 1$ they reproduce eqs.  (\ref{A1}) and
 (\ref{A2}) respectively for $b$ and $c$.
This is consistent with the
fact that a fermionic $bc$ system is completely
 equivalent to a scalar field with a
background charge ${\cal Q} = 1- 2 \lambda$ and with $\epsilon =1$. The 
fields $b$ 
and $c$ can be expressed in terms of the scalar field through 
the bosonization eqs.(\ref{2.25}) and the current
$j(z)$ in eq.(\ref{2.19}) turns out to be the bosonized version
of the fermionic number current in eq.(\ref{2.6a}).
Consequently the zero mode $N$ in eq.(\ref{2.21}) is just the bosonized version
of the fermionic number, as one can see from
\EQ
N = \oint dz j(z) = j_{0}~~.
\label{2.23}
\EN
In the case of a bosonic $bc$ system the central charge of the Virasoro 
algebra in eq.(\ref{Qc}) can be written as:
\beq
c = -1 + 3 {\cal Q}^2 = (1+3 {\cal Q}^2) -2~~,
\label{cc32}
\eeq
that corresponds to the sum of the central charges of a scalar field with 
$\epsilon =-1$ given by $c=1 +3 {\cal Q}^2$ and of a fermionic $bc$ system with
$\lambda =1$ given by $c=-2$. In this case the "bosonization"  rules are given
in eqs.(\ref{bcboso}).
Introducing the new energy momentum tensor as 
\beq
\label{A3}
T(z)=T_\varphi(z)+T_{\eta\xi}(z)~~~~;~~~~
\eeq
where $T_\varphi$ is given in eq. 
(\ref{2.19a}) for $\epsilon=-1$, ${\cal Q}=(-1+2\lambda)$ and 
$T_{\eta\xi}$ is given in eq. (\ref{2.6}) for $\epsilon=1$ and $\lambda=1,$
it is easy to verify that the fields in the r.h.s. of eqs.(\ref{bcboso}) 
have exactly the same conformal weights of a bosonic $(b,c)$ system.
Moreover, if we introduce the sum of $U(1)$ number currents of the scalar
field $\varphi$ and of the fermionic $\xi, \eta$ system:
 \beq
\label{totj}
j(z)=j_\varphi(z)+j_{\eta\xi}(z)=-\partial\varphi(z)+\xi(z)\eta(z)~~,
\eeq
it is easy to verify that the OPE of $j (z)$ with $\beta (z)$ and 
$\gamma (z)$ has no simple pole term implying that 
both $\beta(z)$ and $\gamma(z)$ have 
charge zero with respect to the total $U(1)$  number given by
\beq
\label{totca}
P=(j_0)_\varphi+(j_0)_{\eta\xi}=
\oint \frac{dz}{2\pi \ii }(-\partial\phi +\xi\, \eta)
\eeq
On the other hand the $U(1)$ current for the bosonic $b,c$ system given
in eq.(\ref{2.9}) is instead reproduced in the "bosonized" system by
only the term $(j_0 )_\varphi$ in eq.(\ref{totca}).

\end{document}